\documentclass[acmsmall]{acmart}

\usepackage{multirow}
\usepackage[normalem]{ulem}
\graphicspath{ {./images/} }
\usepackage{hyperref}
\hypersetup{hidelinks,
	colorlinks=true,
	allcolors=black,
	pdfstartview=Fit,
	breaklinks=true}
\usepackage{amsmath}
\usepackage[ruled,linesnumbered]{algorithm2e}

\AtBeginDocument{%
  }

\setcopyright{acmlicensed}
\copyrightyear{2025}
\acmYear{2025}

\begin{document}

\title{Coherence-guided Preference Disentanglement for Cross-domain Recommendations}

\author{Zongyi Xiang}
\email{xzongyi@foxmail.com}
\affiliation{%
  \institution{University of Electronic Science and Technology of China}
  \country{Chengdu, China}
}

\author{Yan Zhang}
\authornote{Corresponding author.}
\email{yan.zhang@cdu.edu.au}
\affiliation{%
  \institution{Faculty of Science and Technology, Charles Darwin University}
  \country{Darwin, Australia}}

\author{Lixin Duan}
\authornotemark[1]
\email{lxduan@uestc.edu.cn}
\affiliation{%
  \institution{Shenzhen Institute for Advanced Study, University of Electronic Science and Technology of China}
  \country{Shenzhen, China}
}

\author{Hongzhi Yin}
\email{h.yin1@uq.edu.au}
\affiliation{%
 \institution{The University of Queensland}
 \streetaddress{4072 St Lucia QLD}
 \country{Brisbane, Australia}}

\author{Ivor W. Tsang}
\email{ivor\_tsang@cfar.a-star.edu.sg}
\affiliation{%
  \institution{CFAR and IHPC, Agency for Science, Technology and Research, Singapore College of Computing and Data Science, Nanyang Technological University}
  \streetaddress{21 Lower Kent Ridge Rd}
  \country{Singapore}}

\renewcommand{\shortauthors}{Z. Xiang et al.}

\begin{abstract}
   Discovering user preferences across different domains is pivotal in cross-domain recommendation systems, particularly when platforms lack comprehensive user-item interactive data. The limited presence of shared users often hampers the effective modeling of common preferences. While leveraging shared items' attributes, such as category and popularity, can enhance cross-domain recommendation performance, the scarcity of shared items between domains has limited research in this area. To address this, we propose a Coherence-guided Preference Disentanglement (CoPD) method aimed at improving cross-domain recommendation by i) explicitly extracting shared item attributes to guide the learning of shared user preferences and ii) disentangling these preferences to identify specific user interests transferred between domains. CoPD introduces coherence constraints on item embeddings of shared and specific domains, aiding in extracting shared attributes. Moreover, it utilizes these attributes to guide the disentanglement of user preferences into separate embeddings for interest and conformity through a popularity-weighted loss. Experiments conducted on real-world datasets demonstrate the superior performance of our proposed CoPD over existing competitive baselines, highlighting its effectiveness in enhancing cross-domain recommendation performance. The code is available at \url{https://github.com/XiangZongyi/CoPD}.
\end{abstract}

\ccsdesc[500]{Information systems~Recommender systems}



\keywords{Cross-domain recommendation, preference disentanglement, graph collaborative filtering, user popularity}


\maketitle

\section{Introduction}
The utilization of e-commerce platforms is steadily increasing as consumers seek the best products to meet their requirements. Recommender systems facilitate precise item selection based on sufficient user-item interactive data, user profiles, and item attributes. For instance, various recommender systems, taking Netflix for movies as an example, try to help users discover relevant movies based on users' past preferred movies. These systems commonly leverage user-item interactive data, including explicit or implicit ratings of users to movies, together with user reviews associated with particular movies to enhance the recommendation accuracy. Due to the limited number of user-item interactions, existing cross-domain recommendations generally transfer user preference information from other domains into a target domain. 

Cross-domain recommendation (CDR) \cite{tang2012cross, DDTCDR,fernandez2012cross, zhu2019dtcdr, DisenCDR,BiTGCF, xu2022metacar} provides a strategy to alleviate the issues caused by sparse user-item interaction data, which generally utilize transfer learning approaches \cite{DDTCDR,fernandez2012cross,hu2022artificial} to transfer users' preference information from a source domain to a target domain by employing their shared users. For example, when a user views a movie, the CDR methods can recommend a book with the same theme as the movie in the Book domain because the CDR methods can capture the shared users' preferences. On the other hand, understanding a user's preference for books facilitates us in recommending related movies.
However, the scarcity of shared users across two domains significantly constrains the performance of cross-domain recommendations~\cite{tang2012cross,DDTCDR}. Tang et al. \cite{tang2012cross} emphasized the challenge of sparse overlapping users between domains, hindering the effective transfer of knowledge. Similarly, Li et al. \cite{li2017collaborative} highlighted domain shift and heterogeneity issues, where differences in item distributions and user behaviors across domains posed obstacles in aligning representations and transferring preferences. 
Fernandez et al.~\cite{fernandez2012cross} noted that the cold-start problem, particularly affecting new user or item entries in cross-domain systems, exacerbating the challenge of accurate recommendations based on limited shared information.
Recently, some dual-target CDR methods have successfully improved the results with sparse overlapping users~\cite{zhu2019dtcdr, DDTCDR, GA-DTCDR, PPGN}. These methods integrate the information of different domains by combining the representations of overlapping users, establishing the connection between domains, and enhancing recommendation performance across multiple domains~\cite{zang2022survey, zhu2021cross}. 
Nevertheless, Zhu et al.~\cite{zhu2019dtcdr} identified a potential loss of domain-specific information during shared preference transfer, risking oversimplification and excluding crucial domain-specific characteristics that could enhance accuracy. To tackle this problem, some dual-target CDR methods aim to disentangle domain-shared and domain-specific information~\cite{DisenCDR, DIDA-CDR}, thereby improving model transfer effectiveness.

\begin{figure}
    \centering
    \includegraphics[width=1.0\linewidth]{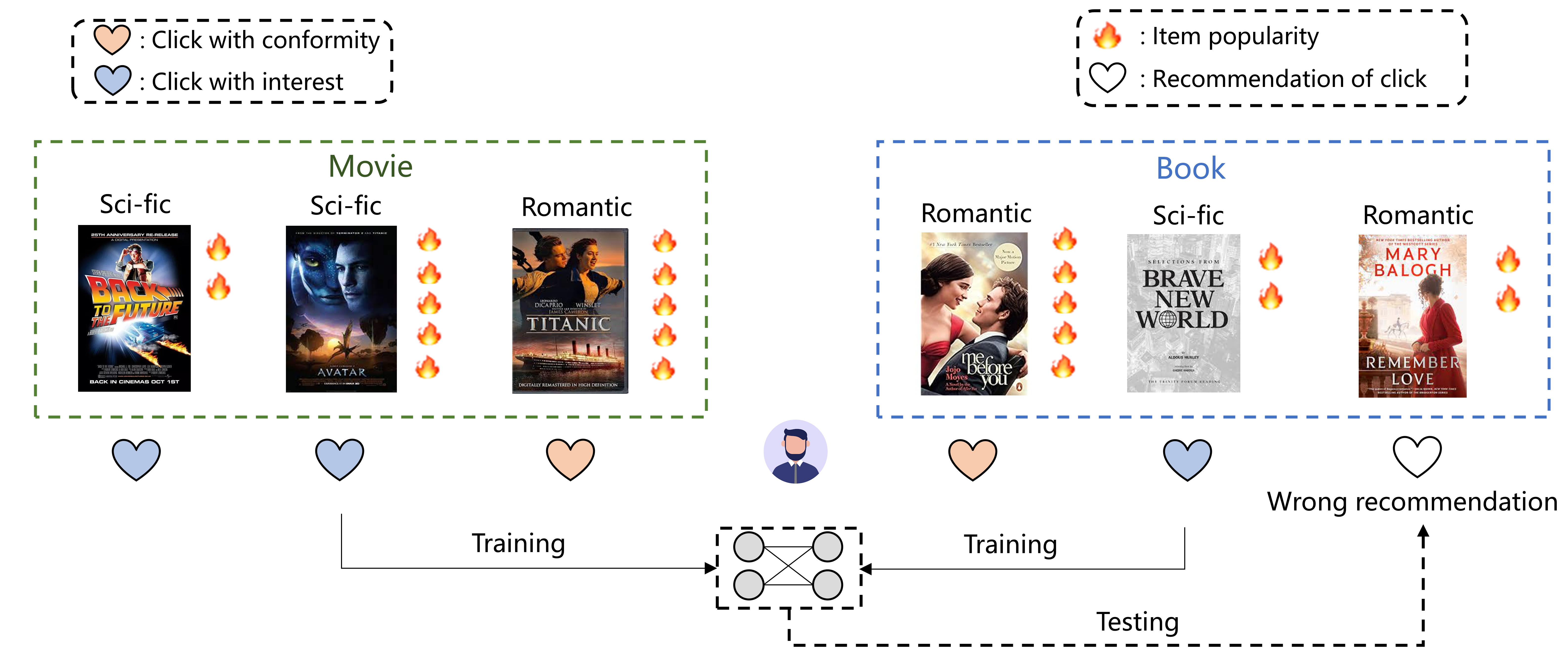}
    \caption{An example of a user's click behavior across two domains. The recommendation model, trained on user preferences from both domains, incorrectly suggests a romantic book. The user clicks on romantic movies/books with his conformity (orange heart-shaped icon) but demonstrates genuine interest by clicking on science fiction movies/books (blue heart-shaped icon).}
    \label{fig:example_user}
\end{figure}

Although the aforementioned dual-target CDR methods achieve promising results on the task with limited overlap among users, they mainly disentangled users' preferences based on domains (e.g., domain-shared and domain-specific), thus neglecting the users' inherent intention towards items. 
Recent research on disentangling users' intents (preferences) in a single domain has achieved more precise recommendations~\cite{DICE, qian2022intent}. This encourages us to investigate the disentanglement of domain-shared users' preferences. 
To this end, we disentangle the shared user preferences into `real interest' and `conformity'.
A user might click on an item due to conformity, disrupting the model's ability to learn the user's real interests. An illustrative example is depicted in Figure~\ref{fig:example_user}, demonstrating a user's clicking behavior across both the Movie and Book domains. The user engages with romantic movies/books out of conformity rather than genuine interest. In fact, the user's true preference lies in scientific movies or books, irrespective of popularity. However, the recommendation model has learned the shared user preferences for science fiction and romance, leading to a recommendation for an unpopular romantic book.

\begin{figure}
    \centering
    \includegraphics[width=0.7\linewidth]{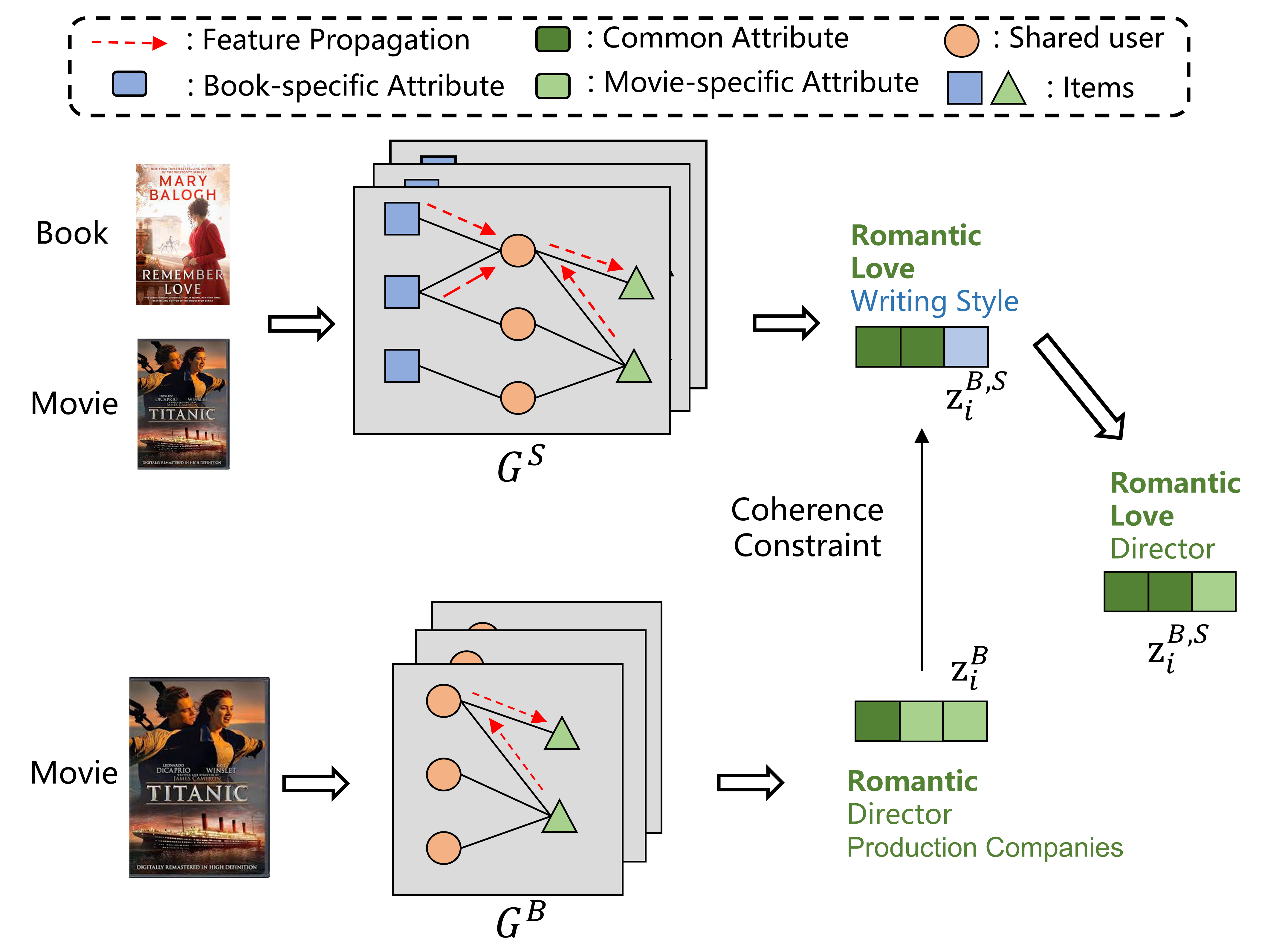}
    \caption{Domain-shared item attributes extracted by coherence constraints. Dark green, blue, and green represent the shared, book-specific and movie-specific attributes, respectively. 
    }
    \label{fig:example_item}
\end{figure}

Disentangling user interest and conformity preferences is challenging due to the sparse overlap among users. Previous dual-target CDR methods primarily emphasized transferring shared user preferences, while neglecting locating which specific preferences are transferred, thereby limiting performance. For example, the recommendation shown in Figure~\ref{fig:example_user} overlooks the user's shared conformity, such as when a user clicks on romantic books and movies due to their popularity rather than romantic attributes. By disentangling users' preferences, we can pinpoint which specific preferences are transferred across two domains. 
Some sequential recommendation methods utilize mixed sequence data~\cite{ye2023dream, xu2023multi} and others construct a Cross-Domain Sequence (CDS) graph that links different domains~\cite{DA-GCN, GA-DTCDR} to obtain domain-shared embeddings. This approach inspired us to acquire domain-shared item attributes using global cross-domain interaction data. Additionally, if shared item attributes (category, popularity, etc.) exist between two domains, the performance of CDR would be greatly improved by transferring both shared user preferences and item attributes. Therefore, we can utilize the learned shared item attributes to guide the learning process of user interest preferences and conformity preferences. Motivated by this intuition, we devise a global-local graph encoder framework to learn the domain-shared and domain-specific embeddings. We aim to learn the shared item attributes through a cross-domain global graph $G^S$, where information can be propagated across both domains. Additionally, we utilize coherence constraints across the two domains to eliminate redundant item attributes. Specifically, Figure~\ref{fig:example_item} illustrates a cross-domain global graph denoted as \(G^{S}\) and a single-domain local graph denoted as \(G^{B}\). In \(G^{S}\), each edge represents an interaction (rating, click, etc.) of a user depicted as an orange circle) with an item in a shared domain \(S\), comprising items from both Book and Movie domains. Meanwhile, edges of \(G^{B}\) denote user-item interactions solely within the Movie domain. Items in the Book domain (A) are denoted as blue rectangles, and items in the Movie domain (B) are denoted as green triangles. For an item $i$ in the Movie domain, we can extract the item embeddings \(z_{i}^{B}\) from the Movie domain and \(z_{i}^{B,S}\) from the shared domain via  Graph Neural Network (GNN), where \(z_{i}^{B,S}\) carries attribute information from both Movie and Book domains through multi-layer information propagation. To keep domain-shared attributes and remove redundant information in $z_{i}^{B,S}$, we enforce coherence constraints between item embeddings \(z_{i}^{B}\) and \(z_{i}^{B,S}\) and attain an enhanced domain-shared item embedding.

This paper proposes a \textbf{Co}herence-guided \textbf{P}reference \textbf{D}isentanglement (CoPD) method to improve cross-domain recommendation performance by extracting shared item attributes and applying them to guide preference disentanglement. Firstly, to learn shared item attributes, we generate the domain-shared item embeddings and enforce coherence constraints on items' domain-specific (source or target domain) and domain-shared embeddings. The coherence constraints in the proposed CoPD emphasize boosting item attribute coherence between a domain-shared item embedding and a domain-specific item embedding. Specifically, Figure~\ref{fig:example_item} shows the coherence constraints imposed on the two embeddings \(z_{i}^{B,S}\) and \(z_{i}^{B}\) is capable of learning an enhanced shared item embedding including domain-shared attributes and domain-specific attributes. Secondly, the learned enhanced shared item embedding guides the preference disentanglement. Specifically, shared user preferences are disentangled as the user's real interest and conformity, and we learn their embeddings respectively by employing a popularity-weighted loss. Finally, we integrate the domain-shared embeddings and domain-specific embeddings to form comprehensive user and item embeddings. The main contributions are summarized as follows:
\begin{itemize}
    \item We propose a Coherence-guided Preference Disentangle (CoPD) framework designed to improve the performance of cross-domain recommendations to tackle the challenge of scarcity shared users between domains. CoPD consists of three components: coherence-guided item attribute alignment, user preference disentanglement guided by the aligned attribute, and embedding fusion and recommendation.
    \item To learn comprehensive item embedding, we propose a global-local graph framework, and add coherence constraints on the embeddings of specific domains and shared domain (the union of two domains) to extract shared item attributes; meanwhile, the comprehensive item embedding keeps essential domain-specific information and drops redundant information in another domain. 
    \item To disentangle shared user preferences into the user's real interest and conformity, we construct a popularity-weighted loss to facilitate this learning process and employ comprehensive item embedding to guide the process of preference disentanglement.
    \item We design experiments to demonstrate the effectiveness of our proposed CoPD. The results show the superior performance of our method over state-of-the-art baselines. Additionally, we conduct ablation studies to evaluate the effectiveness of coherence constraints and preference disentanglement. 
\end{itemize}

\section{Related Work}
\subsection{Cross-Domain Recommendations}
The cross-domain recommendation aims to augment the recommendation effectiveness within the target domain by leveraging insights from other domains. Based on the inference goal of recommendation task, cross-domain recommendation methodologies can be roughly categorized into three categories~\cite{liu2024mcrpl}: single-target domain~\cite{EMCDR, PTUPCDR}, dual-target domain~\cite{BiTGCF, DisenCDR, DIDA-CDR, guo2022reinforcement, DDTCDR, GA-DTCDR}, and multi-target domain~\cite{cui2020herograph, krishnan2020transfer}. Our CoPD belongs to dual-target cross-domain recommendations, focusing on transferring knowledge across two domains to improve recommendation performance in both domains. Dual-target CDR models emphasize identifying overlapping entities across disparate fields and achieving integration by sharing or enhancing the representations of such entities in distinct fields. Some approaches such as CDFM~\cite{CDFM}, employ shallow models as a backbone. CDFM aims to proficiently transfer knowledge by incorporating source domain interactions as supplementary features, based on the principles of Factorization Machines~\cite{FM}. Furthermore, other approaches such as CoNet~\cite{CoNet}, utilize neural networks as a backbone. CoNet facilitates knowledge transfer across domains through the use of cross-connection networks between feed-forward multi-level perceptrons (MLPs). In a different view, DDTCDR~\cite{DDTCDR} pursues the learning of a latent orthogonal mapping function for shared users, enabling the transfer of user preferences across domains. Some methods leverage graph neural networks as a backbone, achieving state-of-the-art performance. For example, PPGN~\cite{PPGN} involves stacking several graph convolution networks on two domains and achieves knowledge transfer by sharing common hidden user features. Similarly, BiTGCF~\cite{BiTGCF} introduced a graph collaborative filtering network, leveraging public users as conduits to facilitate bidirectional cross-domain knowledge transfer, thereby enhancing performance across both domains. DA-GCN~\cite{DA-GCN} and TiDA-GCN~\cite{TiDA-GCN} construct a Cross-Domain Sequence (CDS) graph to model complex associations among users and items across domains, where users and items from each domain are linked as nodes in a graph, with their associations represented as edges. Additionally, they are domain-aware in the sense that they adapt to the specific preferences and behaviors of users within each domain. However, previous methods neglect the disentanglement of domain-shared and domain-specific information, leading to the problem of negative transfer~\cite{Zhou2023FeatureDF, Zhu2021AUF}. Wang et al.~\cite{wang2022disentangled} utilizes the explicit reconstruction loss integrated within the Evidence Lower Bound (ELBO), complemented by additional regularizers such as the disentanglement loss, to learn the desirable disentangled representations. DisenCDR~\cite{DisenCDR} propose a variational bipartite graph encoder (VBGE), and use different VBGEs encode domain-shared and domain-specific representations. Then, DisenCDR learns these representation with two disentanglement regularizers based on mutual information. DIDA-CDR~\cite{DIDA-CDR} captures the domain-shared user representations using an interpolative data augmentation approach and proposes a disentangling module to disentangle domain-specific and domain-independent information to capture comprehensive user preferences.

\subsection{Disentangled Representation Learning}
Disentangled representation learning, initially introduced within the domain of computer vision~\cite{gonzalez2018image}, primarily extracts visual characteristics, including shape, color, and spatial attributes of objects. Beyond its application in computer vision, disentangled representation learning has increasingly found utility within recommender systems in recent years.
Disentangled representation learning in the recommender system focuses on factorizing unobservable structural factors, including domain-based features, item attributes and user preferences toward items. The $\beta$-VAE~\cite{betaVAE} based disentanglement model proposed by Nema et al.~\cite{DisentanglingPR} can identify user tastes by concerning disentangled attributes, enabling users to manipulate recommendations across these attributes. DGCF~\cite{wang2020disentangled} focuses on uncovering users' hidden intentions by disentangling the interaction graph into different intent sub-graphs. It segments the embedding of users or items into multiple blocks corresponding to decoupled intentions. DICE~\cite{DICE} disentangles the embedding representation into specific intents, distinguishing whether users purchase items based on their own interests or blind obedience. DICE established an inverse relationship between interest and popularity, in order to effectively learn interest embeddings. From a different view of conformity, Zhao et al.~\cite{zhao2022popularity} disentangle item attributes into quality (benign conformity) and conformity (harmful conformity), and remove the harmful conformity during inference. Previous approaches mainly segment embeddings into different intent components or assign separate embeddings to users and items to achieve disentanglement. \cite{EDDA, CD2AN, Disen-GNN} validate its capability to disentangle embeddings through a specifically designed disentanglement module. Qian et al.~\cite{qian2022intent} proposes an intent prototype strategy to model popularity and preference intents separately, and learn these intent representations using a novelty-weighted recommendation loss.

\subsection{Differences}
Our CoPD differs significantly from previous studies in several aspects: (1) Before disentangling user preference, we introduce a global-local graph encoder to disentangle domain-shared and domain-specific information. Different from DisenCDR~\cite{DisenCDR} and DIDA-CDR~\cite{DIDA-CDR}, CoPD learns domain-shared user representations through a global graph encoder, allowing information from both domains to influence the user representation. Additionally, we can learn domain-shared item representations using the global graph encoder and two coherence constraint regularizers, which have not been explored in previous dual-target CDR methods. Unlike PPGN~\cite{PPGN}, which constructs a cross-domain user-item interaction graph to learn user and item representations, CoPD acquires domain-specific user representations through a local graph encoder. This approach allows CoPD to address the negative transfer problem and ensure independent learning of domain-specific representations. (2) Our preference disentanglement strategy differs from other disentangled representation learning methods. While previous methods aim to disentangle both user and item embeddings~\cite{wang2020disentangled, DICE}, accomplishing this in dual-target CDR tasks presents challenges due to the data sparsity problem. Hence, CoPD focuses on disentangling user embeddings into interest and conformity embeddings and disentangling user and item embeddings into domain-shared and domain-specific embeddings. The learning of users' interest and conformity embeddings is guided by disentangled domain-shared item embeddings. Specifically, to learn domain-shared user preferences, CoPD disentangles domain-shared user embeddings into domain-shared interest and conformity embeddings, which are guided by domain-shared item attribute embeddings. Furthermore, CoPD employs two preference encoders to disentangle preference embeddings. However, it is difficult to sample negative samples in CDR~\cite{RealHNS}. To facilitate the effective training of these preference encoders, we employ a popularity-weighted Softmax loss, which enables the selection of challenging examples from multiple negative. Based on the causality analysis in DICE~\cite{DICE}, although user interests can indeed align with popular items, it is difficult to learn real interests from popular items. We assert an inverse correlation between interest embeddings and popularity weight to learn users' real interests.

\section{Methods}
\begin{table}[htbp]
\renewcommand{\arraystretch}{1.3}
\centering
\caption{The important notations.}
\label{tab:notation}
\resizebox{1.0\textwidth}{!}{
    \begin{tabular}{cl}
    \hline
        Symbol &  Definition  \\ \hline
        $A$, $B$ & Domain A and Domain B.\\
        $\mathcal{U}$ & The shared user set.\\
        $\mathcal{I}^{A}$, $\mathcal{I}^{B}$ & The item sets in domain A and domain B.\\ 
        $\mathcal{R}^{A}$, $\mathcal{R}^{B}$ & The binary interaction matrices in domain A and domain B. \\
        $G^A, G^B, G^S$ & The graphs constructed by interaction matrices in domain A, domain B and shared domain. \\
        $\mathcal{N}_u$ & The item set interacted by user $u$\\
        $\mathcal{D}_{train}^{A}, \mathcal{D}_{train}^{B}$ & The training datasets in domain A and domain B. \\
        $\mathcal{D}_{test}^{A}, \mathcal{D}_{test}^{B}$ & The test datasets in domain A and domain B. \\
        \hline
        $H_{u}^{A}, H_{u}^{B}, H_{u}^{S}$ & The initialized user embedding matrices in domain A, domain B and shared domain. \\
        $H_{i}^{A}, H_{i}^{B}$ & The initialized item embedding matrices in domain A and domain B.\\
        $h_{u/i}^{*}$ & The sampled embedding from corresponding embedding matrix $H_{u/i}^{*}$. \\
        $z_{u/i}^{A}, z_{u/i}^{B}, z_{u/i}^{S}$ & The user/item graph embeddings propagated by $G^A, G^B$, $G^S$. \\
        $z_{i}^{A,S}, z_{i}^{B,S}$ & The shared item embeddings in domain A and domain B split from $z_{i}^{S}$. \\
        $e_{u/i}^{A}$, $e_{u/i}^{B}$ & The integrated user/item embeddings in domain A and domain B.\\
        \hline
    \end{tabular}
}
\end{table}
\subsection{Preliminaries\label{sec:preliminaries}}
In this section, we first describe the notations used in our paper. We list the important notations of this paper in Table~\ref{tab:notation} for better readability.
For dual-target cross-domain recommendation, there are two domains, A and B, with a shared user set $\mathcal{U}$. The item sets in domain A and domain B are denoted as \(\mathcal{I}^{A}\) and \(\mathcal{I}^{B}\), respectively. Additionally, two binary interaction matrices are defined as \(\mathcal{R}^A \in \{0,1\}^{\vert{\mathcal{U}}\vert \times \vert{\mathcal{I}^A}\vert}\) and \(\mathcal{R}^B \in \{0,1\}^{\vert{\mathcal{U}}\vert \times \vert{\mathcal{I}^B}\vert}\) in domain A and domain B, where $\vert{\mathcal{U}}\vert$ is the size of the user set and $\vert{\mathcal{I}^A}\vert$/$\vert{\mathcal{I}^B}\vert$ are the size of the item sets of domain A and B. 
We embed the user set $\mathcal{U}$ with $H_{u}^{A} \in \mathbb{R}^{\vert{\mathcal{U}} \vert \times d}$, $H_{u}^{B} \in \mathbb{R}^{\vert{\mathcal{U}}\vert \times d}$ and $H_{u}^{S} \in \mathbb{R}^{\vert{\mathcal{U}}\vert \times d}$ as domain-A-specific, domain-B-specific and domain-shared initialized embedding matrices, and \(d\) denotes the embedding dimension. Meanwhile, we embed the item sets with \(H_{i}^{A} \in \mathbb{R}^{\vert{\mathcal{I}^{A}}\vert \times d}\) and \(H_{i}^{B} \in \mathbb{R}^{\vert{\mathcal{I}^{B}}\vert \times d}\) in domain A and domain B respectively. Note that we use uppercase (e.g., $H_{u}^{A}$) to represent entire embedding matrices and lowercase letters (e.g., $h_{u}^{A}$) to represent sampled embeddings from embedding matrices. \par
The domain-based disentanglement and preference-based disentanglement are used in our method. For domain-based disentanglement, we disentangle user embeddings into domain-shared embeddings \(z_{u}^{S}\) and domain-specific embeddings $z_{u}^{A}$/$z_{u}^{B}$. Different from previous methods, we also disentangle item embeddings into domain-shared embeddings $z_{i}^{A,S}$/$z_{i}^{B,S}$ and domain-specific embeddings $z_{i}^{A}$/$z_{i}^{B}$. In our dual-target CDR task, the items in domains A and B do not overlap. Here, $z_{i}^{A,S}$ denotes the item embeddings within domain A, capturing shared attributes across both domains and the same for $z_{i}^{B,S}$. For preference-based disentanglement, we define user interest embeddings as \(z_{u}^{int}\) and user conformity embeddings as \(z_{u}^{conf}\). Finally, we use $e_{u}^{A}$/$e_{u}^{B}$ and $e_{i}^{A}$/$e_{i}^{B}$ to represent the fused user and item embeddings in domain A of B. \par

\subsection{Overall Framework}

\begin{figure}[htbp]
    \centering
    \includegraphics[width=1.0\textwidth]{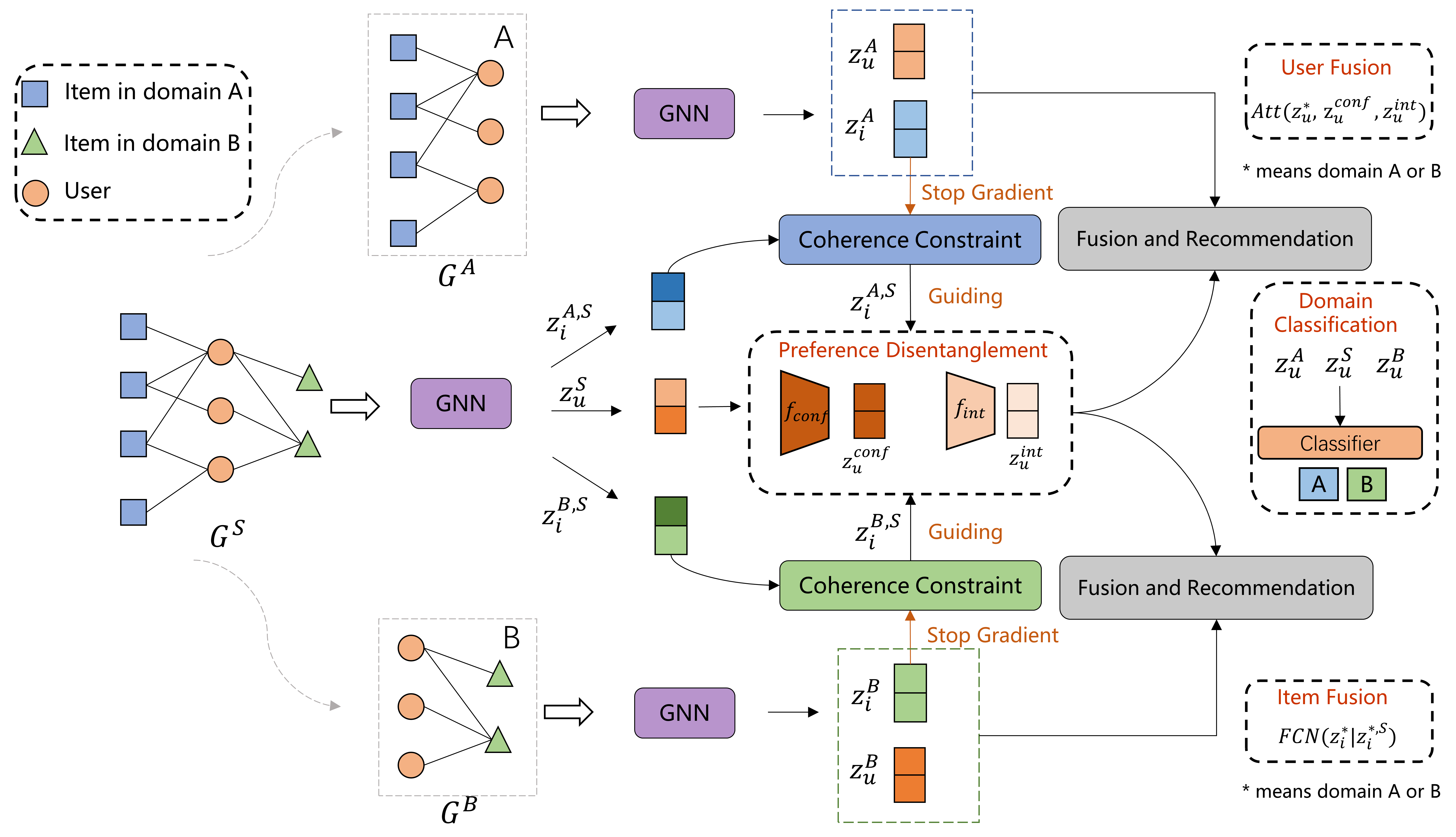}
    \caption{The CoPD framework comprises three primary components: Coherence-guided Item Attribute Alignment, User Preference Disentanglement guided by Aligned Attribute, and Embedding fusion and Recommendation. The process begins by generating graph-enhanced embeddings using three GNN models. Subsequently, we align the item attribute embeddings and disentangle user preference embeddings. Finally, we fuse the domain-shared user/item embeddings with their corresponding domain-specific embeddings to generate comprehensive recommendations.}
    \label{model}
\end{figure}

In this section, we present a novel approach called \textbf{Co}herence-guided \textbf{P}reference \textbf{D}isentanglement, denoted as CoPD, designed for dual target cross-domain recommendations, as depicted in Figure~\ref{model}. 
Firstly, we introduce the graph collaborative filtering backbone in our CoPD to obtain graph embeddings for users and items. The details can be seen in Section~\ref{sec:backbone}.
Next, we introduce CoPD's three components: Coherence-guided Item Attribute Alignment, User Preference Disentanglement, and Embedding Fusion and Recommendation: \par
\textbf{(1) Coherence-guided Item Attribute Alignment.} To acquire domain-shared embeddings for users and items, we create a global graph connecting domain A and domain B. This facilitates the propagation of feature information between the two domains, leveraging overlapping users as connectors. Furthermore, to ensure precise domain-shared item embeddings, we apply coherence constraints that align them with domain-specific item embeddings. Their coherence constraints assist in enabling domain-shared item embeddings to capture shared attributes while filtering out unnecessary ones comprehensively. The details can be seen in Section~\ref{sec:item}. \par
\textbf{(2) User Preference Disentanglement.} We disentangle domain-shared user embeddings into interest embeddings with an interest encoder and conformity embeddings with a conformity encoder. To disentangle user preference effectively, we utilize a popularity-weighted loss function to train interest and conformity encoders. Specifically, we establish a positive correlation between conformity preferences and popularity, and a negative correlation between interest preferences and popularity. In addition, we feed the learned domain-shared item embeddings into the popularity-weighted loss to facilitate the disentanglement of user preferences. The details can be seen in Section~\ref{sec:user}. \par
\textbf{(3) Embedding Fusion and Recommendation.} Both domain-shared and domain-specific information are important for recommendations. To integrate this information for users, we employ an attention mechanism to fuse domain-shared interest and conformity embeddings with domain-specific embeddings. Similarly, for items, we concatenate the domain-shared embeddings with corresponding domain-specific embeddings and feed them into a fully connected network. Finally, we employ the inner product operation for integrated user and item embeddings to make recommendations. The details can be seen in Section~\ref{sec:fusion}.

\subsection{Graph Collaborative Filtering Backbone\label{sec:backbone}}
Graph Neural Network (GNN) based collaborative filtering methods~\cite{LightGCN, NGCF, NCL} fundamentally aggregate neighboring node features to enhance the overall representation for users and items within the graph. Generally, the aggregation scheme is formulated as follows:
\begin{align}
    h_{u}^{(l+1)}=Agg(h_{u}^{(l)}, h_{i}^{(l)}; i \in \mathcal{N}_u),
    \label{eq:agg}
\end{align}
\begin{align}
    z_u = Readout([h_{u}^{(0)}, h_{u}^{(1)}, ..., h_{u}^{(L)}]), 
    \label{eq:readout}
\end{align}
where \(Agg(\cdot)\) represents an aggregation function, such as a weighted sum aggregator or mean aggregator, and $\mathcal{N}_u$ denotes the neighbor set of user $u$ in the interaction matrix. The \(h_{u}^{(l)}\) and \(h_{i}^{(l)}\) denote the refined embeddings for user \(u\) and item \(i\) after \(l\) layers of information propagation, where \(h_{u}^{0}\) and \(h_{i}^{0}\) are sampled from two learnable embedding matrices $H_u$ and $H_i$. The $L$ denotes the number of GNN layers and the readout function $Readout(\cdot)$ further encodes all representations of $[h_{u}^{(0)}, h_{u}^{(1)}, ..., h_{u}^{(L)}]$ to acquire the ultimate user $u$ representations for recommendation. Similarly, the informative representations of items can be acquired analogously. \par
In CoPD, we produce user and item representations by applying GNN-based methods. Following LightGCN~\cite{LightGCN}, we remove the nonlinear activation and feature transformation in Equation~\eqref{eq:agg} as:
\begin{equation}
\begin{aligned}
    h_{u}^{(l+1)}= \sum_{i \in \mathcal{N}_u}{\frac{1}{\sqrt{\vert{\mathcal{N}_u}\vert} \sqrt{\vert{\mathcal{N}_i}\vert}}h_{i}^{(l)}}, \\
    h_{i}^{(l+1)}= \sum_{u \in \mathcal{N}_i}{\frac{1}{\sqrt{\vert{\mathcal{N}_i}\vert} \sqrt{\vert{\mathcal{N}_u}\vert}}h_{u}^{(l)}},
\end{aligned}
\end{equation}
where \(\frac{1}{\sqrt{\vert{\mathcal{N}_u}\vert} \sqrt{\vert{\mathcal{N}_i}\vert}}\) serves as a symmetric normalization term, addressing the concern of embedding scale expansion resulting from graph convolution operations. After the propagation through $L$ layers, we employ the average sum function in Equation~\eqref{eq:readout} to encode the representations generated across all layers and obtain the conclusive representations as follows:
\begin{equation}
\begin{aligned}
    z_{u}=\frac{1}{(L+1)} (h_{u}^{(0)} + h_{u}^{(1)} + ... + h_{u}^{(L)}), \\
    z_{i}=\frac{1}{(L+1)} (h_{i}^{(0)} + h_{i}^{(1)} + ... + h_{i}^{(L)}),
\end{aligned}
\end{equation} \par

\subsection{Coherence-guided Item Attribute Alignment\label{sec:item}}
To acquire domain-shared and domain-specific representations, we create global and local graphs using the corresponding user-item interaction matrices. Specifically, we create a global interaction matrix \(\mathcal{R}^S\), where \(\mathcal{R}^S \in \{0,1\}^{\vert{\mathcal{U}}\vert \times (\vert{\mathcal{I}^A}\vert + \vert{\mathcal{I}^B}\vert)}\) is formed by concatenating \(\mathcal{R}^A\) and \(\mathcal{R}^B\) (\(\mathcal{R}^S\)=\(\mathcal{R}^A\) $\Vert$ \(\mathcal{R}^B\)). The construction of global and local graphs can be written as follows:
\begin{align}
     G^* = \left(                 
        \begin{array}{cc}   
            0 & \mathcal{R^*} \\  
            \mathcal{R^*}^\top & 0 \\  
        \end{array}
        \right), 
\end{align}
where $G^*$ ($G^A, G^B$ or $G^S$) is constructed by corresponding interaction matrices $\mathcal{R}^*$ ($\mathcal{R}^A, \mathcal{R}^B$ or $\mathcal{R}^S$).\par
After that, we denote the GNN backbone as $f_{gnn}$ and domain-shared and domain-specific embeddings of user/item can be acquired by:
\begin{equation}
    z_{u}^{S}, z_{i}^{S}=f_{gnn}(h_{u}^{S}, h_{i}^{A}\Vert h_{i}^{B}, G^S), 
    \label{eq:f_gnn_shared}
\end{equation}
\begin{equation}
\begin{aligned}
    z_{u}^{A}, z_{i}^{A}=f_{gnn}(h_{u}^A, h_{i}^{A}, G^A), \\
    z_{u}^{B}, z_{i}^{B}=f_{gnn}(h_{u}^B, h_{i}^{B}, G^B),
    \label{eq:f_gnn_single}
\end{aligned}
\end{equation}
please note that we concatenate the origin item embeddings $h_{i}^{A}\Vert h_{i}^{B} \in \mathbb{R}^{(\vert \mathcal{I}^{A} \vert + \vert \mathcal{I}^{B} \vert) \times d}$ to facilitate enhanced information propagation for items across the two domains. After the information aggregation and readout of GNN, we obtain the domain-shared item embeddings $z_{i}^{S}$. We then split $z_{i}^{S}$ into $z_{i}^{A,S}$ for domain A and $z_{i}^{B,S}$ for domain B. \par
The domain-shared attributes \(z_{i}^{A,S}\) and \(z_{i}^{B,S}\) can by propagated by the common user. However, the domain-shared item embeddings may incorporate redundant domain-specific attributes, which could be detrimental for recommendations in both domains. To solve this problem, our CoPD introduces coherence constraints to maintain alignment between the domain-shared item embeddings (\(z_{i}^{A,S}\) and \(z_{i}^{B,S}\)) and the corresponding domain-specific item embeddings (\(z_{i}^{A}\) and \(z_{i}^{B}\)). 
Furthermore, taking domain A as an example, we freeze the parameters of \(z_{i}^{A}\) and introduce coherence constraints to \(z_{i}^{A,S}\) and \(z_{i}^{A}\). This process aims to eliminate irrelevant features from domain B in \(z_{i}^{A,S}\). Specifically, we utilize cosine similarity to enforce coherence constraints between domain-shared and domain-specific embeddings. Cosine similarity is chosen because it can capture both magnitude and angle similarity. We consider the item \(i\) clicked by the user \(u\) (\(i\in \mathcal{N}_u\)) as the positive sample. In contrast, the item \(j\) not clicked by user \(u\) (\(j \notin \mathcal{N}_u\)) as the negative sample. These constraints are applied to both domain A and domain B. Taking domain A as an example, the proposed coherence constraints learning objective is to minimize the following function:      
\begin{equation}
    \mathcal{L}_{cc}^{A}=1 - \frac{z_{i}^{A,S} \cdot SG(z_{i}^{A})}{\Vert z_{i}^{A,S} \Vert \Vert z_{i}^{A} \Vert} + \max (0, \frac{z_{i}^{A,S} \cdot SG({z}_{j}^{A})}{\Vert z_{i}^{A,S} \Vert \Vert z_{j}^{A} \Vert}), \\
\end{equation}
where $SG(\cdot)$ denotes the stop gradient operation. Specifically, the stop-gradient facilitates the independent learning of item attributes and prevents extraneous noise inherited from domain-shared item embedding. Similarly, the learning objective of domain B can be acquired analogously.\par

As mentioned above, we utilize coherence constraints to remove the redundant information in domain-shared item embeddings by aligning them with domain-specific item representations. Specifically, we designate the domain-specific embeddings as the anchor and adjust the domain-shared embeddings to align with this anchor. 
The objective of the coherence constraints is formally expressed as follows:
\begin{equation}
    \mathcal{L}_{cc} = \mathcal{L}_{cc}^{A} + \mathcal{L}_{cc}^{B}.
    \label{eq:loss_ac}
\end{equation}

\subsection{User Preference Disentanglement Guided by Aligned Item Attributes\label{sec:user}}
This section focuses on disentangling domain-shared user embeddings into interest and conformity embeddings with the assistance of domain-shared item embeddings. Firstly, motivated by \cite{recguru, Meta-FDMixup}, we introduce a domain classifier denoted as $f_{dom}(\cdot)$ to ensure that the GNN backbone learns domain-shared and domain-specific user preferences.
This classifier incorporates a single fully connected layer to predict the domain probability associated with user embeddings.
Following Meta-FDMixup~\cite{Meta-FDMixup}, we define the ground truth for domain-specific user embeddings as \(y_A\) and \(y_B\) for their respective domains. The ground truth for domain-shared user embeddings is denoted as \(y_S\). We hope that the learned domain-specific user embeddings (e.g., \(z_{u}^{A}\) and \(z_{u}^{B}\)) can be easily categorized as a particular domain and the learned domain-shared user embeddings \(z_{u}^{S}\) cannot be classified into either domain A or domain B. Formally, let one and zero denote the label of domain A and domain B, respectively. All values in \(y_A\) are set as one, and all in \(y_B\) are set as zero. In particular, all entries in \(y_S\) are set as 0.5, because \(z_{u}^{S}\) contains common user preferences information across both domain A and domain B. The objective of the domain classifier is formulated as follows,
\begin{equation}
\begin{aligned}
    \mathcal{L}_{dom}=&\frac{1}{3} \sum KL(f_{dom}(z_{u}^{S}),y_S)\nonumber\\
        & +CE(f_{dom}(z_{u}^{A}),y_A)+CE(f_{dom}(z_{u}^{B}),y_B),
\end{aligned}
\end{equation}
where the \(KL(\cdot, \cdot)\) and the \(CE(\cdot,\cdot)\) denote the Kullback-Leibler divergence loss and cross-entropy loss, respectively. The domain classifier assists in capturing domain-shared user preferences for \(z_{u}^{S}\). \par
Subsequently, the learned domain-shared user embedding is fed into a preference extractor to extract the domain-shared conformity and interest preferences. We implemented our preference extractor with two encoders, i.e., conformity encoder denoted as $f_{conf}$ and interest encoder denoted as $f_{int}$. For simplicity, both of them are implemented as two-layer fully connected layers. We then disentangle the domain-shared user embeddings into two preference embeddings based on the outputs of these two preference encoders:
\begin{equation}
    \begin{aligned}
        z_{u}^{int}=f_{int}(z_{u}^{S}) \\
        z_{u}^{conf}=f_{conf}(z_{u}^{S})
    \end{aligned}
    \label{eq:disentangle}
\end{equation}
where \(z_{u}^{int}\) representing interest embeddings and \(z_{u}^{conf}\) representing conformity embeddings.\par 
After obtaining the disentangled preference embeddings, we hope $z_{u}^{int}$ and $z_{u}^{conf}$ reflect the preferences of user interest and conformity. We introduced the domain-shared item embeddings as mentioned in Section~\ref{sec:item} (i.e., $z_{i}^{A,S}$ and $z_{i}^{B,S}$), which encompasses common attributes for domain A and domain B, to facilitate the representation learning for disentangled preference embeddings. Bayesian Personalized Ranking(BPR) loss~\cite{BPRMF} and Softmax loss~\cite{rendle2021item} can be used to measure the quality of interaction predictions. We employ the Softmax loss here as it inherently facilitates the selection of challenging examples from multiple negative~\cite{lian2020personalized, yi2019sampling, zhang2022incorporating, zhang2023invariant}, thus amplifying the process of learning particular user preference. 
There are two Softmax losses (i.e., conformity loss $\mathcal{L}_{conf}$ and interest loss $\mathcal{L}_{int}$) for learning the interest and conformity preferences. However, if the losses are simply summed without distinction for different items, the leaned representation $z_{u}^{int}$ and $z_{u}^{conf}$ cannot accurately reflect the preferences for user interest and conformity over items. Therefore, we integrate item popularity signal as a weight into the Softmax losses. Item popularity $p_{i}$ is defined as the number of historical interactions that an item $i$ has been involved in.  A more popular item $i$ has a higher item popularity $p_{i}$. We then normalize $p_{i}$ to a range of 0 to 1 using min-max normalization. We aim to establish a positive correlation between the conformity loss $\mathcal{L}_{conf}$ and the popularity weight $p_{i}$. For highly popular items, the weight $p_{i}$ in the conformity $\mathcal{L}_{conf}$ would be close to 1, ensuring that popular items are primarily attributed to user's conformity preference $z_{u}^{conf}$. Conversely, we aim to establish a negative correlation between the interest loss $\mathcal{L}_{int}$ and the popularity weight $p_{i}$. We use $1-p_i$ as the weight of $\mathcal{L}_{int}$. For less popular items, the weight $1-p_i$ would be close to 1, allowing the model to capture the users' interest preference $z_{u}^{int}$ for certain item attributes. For more popular items, the weight would be close to 0 , thereby reducing the influence of item popularity and enabling the model to learn the user's genuine interests. Taking domain A as an example, popularity weight in Softmax loss can be defined as $p_{i}^{A}$. We minimize the popularity-weighted Softmax losses for representative learning:
\begin{align}
    \mathcal{L}^{A}_{conf}=-\sum_{(u,i)\in \mathcal{D}_{train}^A} p_{i}^{A} \log \frac {exp(z_{u}^{conf} \cdot z_{i}^{A,S}/ \tau)}{\sum_{k \in \mathcal{I}^A} exp(z_{u}^{conf} \cdot z_{k}^{A,S}/ \tau)},
    \label{eq:conf}
\end{align}
\begin{align}
    \mathcal{L}^{A}_{int}=-\sum_{(u,i)\in \mathcal{D}_{train}^A} (1-p_{i}^{A}) \log \frac{exp(z_{u}^{int} \cdot z_{i}^{A,S}/ \tau)}{\sum_{k \in \mathcal{I}^A} exp(z_{u}^{int} \cdot z_{k}^{A,S}/ \tau)}
    \label{eq:int},
\end{align}
where \(\tau\) represents the temperature hyper-parameter of softmax~\cite{wang2021understanding} and $\mathcal{D}_{train}^A$ represents the training set in domain A. 
Similarly, we use $p_{i}^{B}$ to represent item popularity in domain B and $\mathcal{D}_{train}^B$ represents the training set in domain B. The weighted Softmax losses in domain B can be calculated as follows:
\begin{align}
    \mathcal{L}^{B}_{conf}=-\sum_{(u,i)\in \mathcal{D}_{train}^B} p_{i}^{B} \log \frac {exp(z_{u}^{conf} \cdot z_{i}^{B,S}/ \tau)}{\sum_{k \in \mathcal{I}^B} exp(z_{u}^{conf} \cdot z_{k}^{B,S}/ \tau)},
    \label{eq:conf_B}
\end{align}
\begin{align}
    \mathcal{L}^{B}_{int}=-\sum_{(u,i)\in \mathcal{D}_{train}^B} (1-p_{i}^{B}) \log \frac{exp(z_{u}^{int} \cdot z_{i}^{B,S}/ \tau)}{\sum_{k \in \mathcal{I}^B} exp(z_{u}^{int} \cdot z_{k}^{B,S}/ \tau)}
    \label{eq:int_B},
\end{align}
We continue to utilize $z_{u}^{int}$ and $z_{u}^{conf}$ in Equations~\eqref{eq:conf_B} and Equations~\eqref{eq:int_B} as user embeddings in domain B. We intend these embeddings to extract common interest and conformity preferences shared across domains A and B. Finally, the objective of user preference disentanglement is formulated as follows:
\begin{equation}
\mathcal{L}_{pd}=\mathcal{L}^{A}_{conf} + \mathcal{L}^{A}_{int} + \mathcal{L}^{B}_{conf} + \mathcal{L}^{B}_{int}.
\label{eq:loss_pd}
\end{equation}

\begin{algorithm}
    \SetAlgoLined
    \KwIn{Training interaction data $\mathcal{D}_{train}^{A}$ and $\mathcal{D}_{train}^{B}$, test interaction data $\mathcal{D}_{test}^{A}$ and $\mathcal{D}_{test}^{B}$; Constructed graph $G^A, G^B, G^S$ of domain A, domain B and shared domain; The number of epochs, batch size, learning rate, and the hyper-parameters of CoPD.}
    \KwOut{The learned user/item embedding matrices, the parameters of two encoders for disentangling; The predicted score between user and item for recommendation.}
    Initialize the embedding matrices of users ($H_{u}^{A}, H_{u}^{B}, H_{u}^{S}$) and items ($H_{i}^{A}, H_{i}^{B}$);\\
    \tcp{Training Procedure}
    \For {each epoch} {
        \For {($u, i^A, i^B$) in ($\mathcal{D}_{train}^{A}$ and $\mathcal{D}_{train}^{B}$)} {
         Sample the origin embeddings from initialized embedding matrices;\\
         Obtain the graph embeddings in domain A and domain B according to Equation~\eqref{eq:f_gnn_single}; \\
         Obtain the graph embeddings in shared domain according to Equation~\eqref{eq:f_gnn_shared}, and split the domain-shared item embeddings; \\
         Enforce coherence constraints on domain-shared item embedding according to Equation~\eqref{eq:loss_ac}; \\
         Disentangle domain-shared user embedding according to Equation~\eqref{eq:disentangle} and calculate the preference disentanglement loss according to Equation~\eqref{eq:loss_pd}; \\
         Generate the final representation according to Equation~\eqref{eq:fusion_item} and \eqref{eq:fusion_user}; \\
         Calculate the BPR loss according to Equation~\eqref{eq:loss_rec}; \\
         Calculate the overall loss and optimize the parameters of the entire model according to Equation~\eqref{eq:final_loss}. \\
        }
    }
    \tcp{Inference Procedure}
    \For {u=\{1,2,3,...,$\mathcal{U}\}$} {
        \For {($u, i^A, i^B$) in ($\mathcal{D}_{test}^{A}$ and $\mathcal{D}_{test}^{B}$)} {
         Sample the origin embeddings from initialized embedding matrices; \\
         Obtain the graph embeddings in domain A and domain B according to Equation~\eqref{eq:f_gnn_single}; \\
         Obtain the graph embeddings in shared domain according to Equation~\eqref{eq:f_gnn_shared}, and split the domain-shared item embeddings; \\
         Disentangle domain-shared user embedding according to Equation~\eqref{eq:disentangle}; \\
         Generate the final representation according to Equation~\eqref{eq:fusion_item} and \eqref{eq:fusion_user}; \\
         Calculate the recommendation score according to Equation~\eqref{eq:cal_score}.\\
        }
    }
    \caption{The CoPD Algorithm.}
    \label{alg:CoPD}
\end{algorithm}

\subsection{Embedding fusion and Recommendation\label{sec:fusion}}
In this section, we introduce the fusion mechanism for combining domain-specific and domain-shared features. Additionally, we introduce the prediction layer to make recommendations based on the fused features. From Section~\ref{sec:item} and Section~\ref{sec:user}, we have obtained the learned domain-specific embeddings of users (i.e., \(z_{u}^{A}\) and \(z_{u}^{B}\)) and items (i.e., \(z_{i}^{A}\) and \(z_{i}^{B}\)), along with the learned domain-shared embeddings of the users (i.e., \(z_{u}^{conf}\) and \(z_{u}^{int}\)) and the items (i.e., \(z_{i}^{A,S}\) and \(z_{i}^{B,S}\)). These embeddings enables us to capture both domain-specific and domain-shared preferences, facilitating more accurate and effective recommendations across domains. Taking domain A as an example, we emphasize the importance of both the domain-specific embeddings \(z_{i}^{A}\) and the domain-shared embeddings \(z_{i}^{A,S}\). Each of these embeddings contains distinct yet complementary portions of item attributes. To aggregate the item attributes in domains A and B, we concatenate the domain-shared embeddings with the corresponding domain-specific embeddings. Subsequently, we use one-layer fully connected layers, denoted as $FCN_{A}$ and $FCN_{B}$ ($\mathcal{R}^{2d} \rightarrow \mathcal{R}^{d}$), to fuse the information of embeddings. This transformation allows for effectively integrating domain-specific and domain-shared attributes, resulting in a comprehensive representation of items in each domain. Taking domain A as an example:
\begin{equation}
\begin{aligned}
    e_{i}^{A} = FCN_{A}(z_{i}^{A}\Vert z_{i}^{A,S}),
\end{aligned}
\label{eq:fusion_item}
\end{equation}
where \(\Vert\) denotes the concatenation operation. Users' preferences for interest and conformity can vary across different scenarios and domains. For example, a user might be open to all popular movies in the Movie domain but have a unique interest in music in the Music domain. Additionally, users may exhibit preferences for items based on either learned domain-shared preferences or learned domain-specific preferences. To adapt users' multiple preferences towards different domains, we employ domain-aware attention functions~\cite{DA-GCN, GA-DTCDR} denoted as $Att_{A}$ and $Att_{B}$ ($\mathcal{R}^{3d} \rightarrow \mathcal{R}^{d}$). The attention function aggregates individual preference representations into comprehensive user representations, enabling the incorporation of diverse user preferences. By attentively weighing the contributions of domain-specific, conformity and interest embeddings, the function effectively captures the refinement preferences of users across different domains and scenarios. The attention function in domain A can be written as follows:
\begin{equation}
\begin{aligned}
    e_{u}^{A} = Att_{A}(z_{u}^{A}, z_{u}^{conf}, z_{u}^{int}),
\end{aligned}
\label{eq:fusion_user}
\end{equation} \par
After obtaining the final embeddings of the user and item, we employ the inner product operation to predict the probability that a user $u$ interacts with an item $i$:
\begin{equation}
    \begin{aligned}
        \hat{y}_{ui}^{A}=(e_{u}^{A})^{T}\cdot e_{i}^{A},
    \end{aligned}
    \label{eq:cal_score}
\end{equation}
To learn representations directly from interactions, we employ the BPR loss~\cite{BPRMF} for the recommendation task. The BPR loss is a pairwise loss function that encourages the model to predict observed user-item interactions to have higher scores compared to unobserved interactions:
\begin{equation}
\begin{aligned}
    \mathcal{L}_{rec}^{A} = -\sum\limits_{u=1}^{\vert \mathcal{U} \vert}\sum\limits_{i\in \mathcal{N}_u}\sum\limits_{j\notin \mathcal{N}_u} \ln \sigma(\hat{y}_{ui}^{A}-\hat{y}_{uj}^{A}),
    \label{eq:bpr}
\end{aligned}
\end{equation} 
By minimizing the BPR loss in Equation~\eqref{eq:bpr}, the positive pairs ($u$ and $i$) have higher rating scores than the negative pairs ($u$ and $j$). Similarly, we perform the same optimization function in domain B. The recommendation loss can be summarized as follows:
\begin{align}
    \mathcal{L}_{rec} = \mathcal{L}_{rec}^{A} + \mathcal{L}_{rec}^{B} + \beta \Vert \Theta \Vert^{2},
    \label{eq:loss_rec}
\end{align}
where \(\beta\) controls the \(L_2\) regularization of the GNN backbone parameters $\Theta$.
Finally, we utilize a multi-task learning mechanism to train our CoPD by minimizing the following cross-domain objectives:
\begin{equation}
\mathcal{L}=\mathcal{L}_{rec}+\mathcal{L}_{dom}+\lambda_{1}\mathcal{L}_{cc}+\lambda_{2}\mathcal{L}_{pd},
\label{eq:final_loss}
\end{equation}
where \(\lambda_1\) and \(\lambda_2\) are hyper-parameters to balance the weights assigned to the coherence constraints loss and preference disentanglement loss, respectively. The complete workflow of our CoPD is shown in Algorithm~\ref{alg:CoPD}.

\section{Experiments}
In this work, we claim that (1) the global graph encoder and coherence constraints can help extract domain-shared attributes for item embedding and (2) our CoPD can disentangle users' interest and conformity embeddings guided by domain-shared item embedding. To verify the effectiveness of our CoPD, we conduct extensive experiments on six cross-domain recommendation tasks. Specifically, we provide comprehensive analyses from the subsequent perspectives:
\begin{itemize}
    \item \textbf{RQ1:} Does our method outperforms the state-of-the-art cross-domain baselines?
    \item \textbf{RQ2:} Are the aligned item embeddings learned by the coherence constraints effective for recommendation performance?
    \item \textbf{RQ3:} Does disentangling user interest and conformity embeddings enhance recommendation performance compared to solely disentangling user-shared embedding? Moreover, to what degree does the disentanglement of user interest and conformity embeddings contribute to overall performance?
    \item \textbf{RQ4:} Does our proposed CoPD successfully achieve the desired disentanglement for user preferences? 
    \item \textbf{RQ5:} How do different settings of hyper-parameters influence the performance of CoPD?
\end{itemize}

\begin{table}[htbp]
\renewcommand{\arraystretch}{1.3}
\centering
\caption{Statistics of six dual-target CDR tasks.}
\label{tab:dataset}
\resizebox{0.6\linewidth}{!}{
    \begin{tabular}{ccccc}
        \toprule
        \midrule 
        Tasks & \#Users & \#Items & \#Interactions & \#Density ($\%$) \\ \midrule
        Amazon-Elec & 3325 & 38706 & 118114 & 0.091 \\ 
        Amazon-Phone & 3325 & 17709 & 52966 & 0.089 \\ \midrule
        Amazon-Sport & 9928 & 30796 & 100938 & 0.033 \\
        Amazon-Cloth & 9928 & 39008 & 95369 & 0.024 \\ \midrule
        Amazon-Sport & 4998 & 20845 & 54256 & 0.052 \\
        Amazon-Phone & 4998 & 13655 & 46445 & 0.068 \\ \midrule
        Amazon-Elec & 15761 & 51447 & 224689 & 0.027 \\
        Amazon-Cloth & 15761 & 48781 & 133609 & 0.017 \\ \midrule
        Douban-Movie & 2106 & 9555 & 907219 & 4.508 \\
        Douban-Book & 2106 & 6777 & 95974 & 0.672 \\ \midrule
        Douban-Movie & 1666 & 9555 &  781288 & 4.908 \\
        Douban-Music & 1666 & 5567 & 69681 & 0.751 \\ \midrule
        \bottomrule
        \end{tabular}
}
\end{table}
\subsection{Datasets}
To evaluate the recommendation performance of our proposed CoPD and ensure a fair comparison with previous methods, we evaluate our model on four publicly available Amazon datasets\footnote{\href{http://jmcauley.ucsd.edu/data/amazon/}{http://jmcauley.ucsd.edu/data/amazon/}}~\cite{He2016UpsAD, McAuley2015ImageBasedRO} and three Douban datasets\cite{GA-DTCDR}. The four Amazon datasets are Electronics (\textbf{Elec}), Cell Phones (\textbf{Phone}), Accessories, Sports and Outdoors (\textbf{Sport}) and Clothing, Shoes and Jewelry (\textbf{Cloth}). Following the method introduced in \cite{BiTGCF}, we convert the explicit ratings into implicit feedback, representing ratings as 0 and 1 to indicate whether the user has rated the item. Next, we filter out users with fewer than 5 interactions and items with fewer than 10 interactions. The users in our dual-target cross-domain recommendation tasks are all overlapped, meaning the users at least click one item in each domain; thus, we extract overlapping users in both domains. Following the experimental setting in DisenCDR~\cite{DisenCDR}, we exclude cold-start items from the test set. Finally, we divide the aforementioned four Amazon datasets into four dual-target Cross-Domain Recommendation (CDR) tasks, comprising two tasks with high domain similarity (Elec \& Phone, Sport \& Cloth) and two tasks with lower domain similarity (Sport \& Phone, Elec \& Cloth). The three Douban datasets are Douban-Movie, Douban-Book and Douban-Music. Following the method introduced in GA-DTCDR~\cite{GA-DTCDR} and DIDA-CDR~\cite{DIDA-CDR}. We convert the explicit ratings into implicit feedback. Next, we filter out users and items with fewer than 5 interactions. We also filter out of the cold-start item entry in the test set. Finally, we construct the three Douban datasets for two dual-target CDR tasks. The details of these six dual-target CDR tasks are presented in Table~\ref{tab:dataset}.

\subsection{Evaluation Metrics}
In item recommendation, the Leave-One-Out (LOO) evaluation~\cite{he2016fast} is commonly employed to demonstrate performance~\cite{DisenCDR,BiTGCF, CoNet,NeuralCF, wu2023graph}. To compare the performance of CoPD and other baseline models, we adopt the LOO evaluation method. To construct a test set of each user, one item is preserved as ground truth, and we randomly sample 999 items that the user has not interacted with as negative items. We utilize two widely adopted evaluation metrics: Hit Ratio (HR) and Normalized Discounted Cumulative Gain (NDCG)~\cite{CumulatedGE}. The predicted rank list is cut off at \(K\) = 10. The HR measures whether the test items are on the top-10 list, and the NDCG estimates ranking quality by assigning higher scores to hits at the top ranks. In both metrics, a higher value indicates superior recommendation performance. 

\begin{table}[htbp]
\renewcommand{\arraystretch}{1.3}
\centering
\caption{The overall performance comparison of CoPD and baselines on Amazon CDR tasks, with the best performance highlighted in bold and the second-best performance underlined.}
\label{tab:experiment-amazon}
\resizebox{\textwidth}{!}{
    \begin{tabular}{ccc|cc|cc|cc}
        \toprule
        \midrule 
        \multirow{2}{*}{Similar tasks} 
            & \multicolumn{2}{c}{Amazon-Elec} & \multicolumn{2}{c|}{Amazon-Phone} & \multicolumn{2}{c}{Amazon-Sport} & \multicolumn{2}{c}{Amazon-Cloth} \\ \cmidrule{2-9} 
            & HR@10 & NDCG@10 & HR@10 & NDCG@10 & HR@10 & NDCG@10 & HR@10 & NDCG@10 \\ \midrule
        NeuMF & 0.1617 & 0.0924 & 0.1584 & 0.0802 & 0.1074 & 0.0546 & 0.1118 & 0.0602 \\
        NGCF & 0.1855 & 0.1087 & 0.2279 & 0.1238 & 0.1313 & 0.0687 & 0.1322 & 0.0697 \\
        LightGCN & 0.1917 & 0.1028 & 0.2325 & 0.1272 & 0.1319 & 0.0694 & 0.1358 & 0.0729 \\ \midrule
        DGCF & 0.1969 & 0.1169 & 0.2489 & 0.1488 & 0.1477 & 0.0887 & 0.1114 & 0.0680 \\
        DICE & 0.1012 & 0.0619 & 0.2200 & 0.1309 & 0.1451 & 0.0862 & 0.1072 & 0.0631 \\
        \midrule
        CDFM & 0.1824 & 0.1092 & 0.1797 & 0.0972 & 0.1161 & 0.0633 & 0.1232 & 0.0705 \\
        CoNet & 0.1722 & 0.0986 & 0.1766 & 0.0930 & 0.1209 & 0.0641 & 0.1240 & 0.0662 \\
        DDTCDR & 0.1847 & 0.1108 & 0.1723 & 0.0858 & 0.1186 & 0.0637 & 0.1254 & 0.0713 \\
        PPGN & 0.2168 & 0.1163 & 0.2454 & 0.1334 & 0.1510 & 0.0803 & 0.1423 & 0.0768 \\
        BiTGCF & 0.2214 & 0.1220 & 0.2571 & 0.1393 & 0.1483 & 0.0795 & 0.1468 & 0.0793 \\ \midrule
        DisenCDR & \uline{0.2457} & \uline{0.1451} & \uline{0.2876} & \uline{0.1613} & 0.1755 & 0.0946 & 0.1631 & 0.0903 \\
        DIDA-CDR & 0.2121 & 0.1252 & 0.2517 & 0.1394 & \uline{0.1890} & 0.1080 & \uline{0.2001} & \uline{0.1148} \\ 
        ETL & 0.2172 & 0.1302 & 0.2712 & 0.1496 & 0.1611 & 0.0939 & 0.1471 & 0.0885 \\
        UniCDR & 0.2292 & 0.1383 & 0.2472 & 0.1377 & 0.1837 & \uline{0.1098} & 0.1785 & 0.1120 \\ \midrule
        CoPD & \textbf{0.2742} & \textbf{0.1653} & \textbf{0.3275} & \textbf{0.1998} & \textbf{0.2594} & \textbf{0.1649} & \textbf{0.2247} & \textbf{0.1495} \\  \midrule
    \end{tabular}
}
        \\ \hspace* {1ex} \\ 
\resizebox{\textwidth}{!}{
    \begin{tabular}{ccc|cc|cc|cc}
        \midrule 
        \multirow{2}{*}{Dissimilar tasks} 
            & \multicolumn{2}{c}{Amazon-Sport} & \multicolumn{2}{c|}{Amazon-Phone} & \multicolumn{2}{c}{Amazon-Elec} & \multicolumn{2}{c}{Amazon-Cloth} \\ \cmidrule{2-9} 
            & HR@10 & NDCG@10 & HR@10 & NDCG@10 & HR@10 & NDCG@10 & HR@10 & NDCG@10 \\ \midrule
        NeuMF & 0.1011 & 0.0519 & 0.1467 & 0.0780 & 0.2008 & 0.1179 & 0.1084 & 0.0580 \\
        NGCF & 0.1606 & 0.0853 & 0.1707 & 0.0922 & 0.2020 & 0.1174 & 0.1086 & 0.0580 \\
        LightGCN & 0.1633 & 0.0916 & 0.1647 & 0.0895 & 0.1997 & 0.1073 & 0.1124 & 0.0611 \\ \midrule
        DGCF & 0.1882 & 0.1147 & 0.1990 & 0.1182 & 0.1645 & 0.0998 & 0.0924 & 0.0555 \\
        DICE & 0.1531 & 0.0901 & 0.1980 & 0.1131 & 0.1814 & 0.1145 & 0.0980 & 0.0554 \\
        \midrule
        CDFM & 0.1197 & 0.0655 & 0.1632 & 0.0901 & 0.2109 & 0.1189 & 0.1037 & 0.0563 \\
        CoNet & 0.1288 & 0.0691 & 0.1660 & 0.0915 & 0.2126 & 0.1261 & 0.1135 & 0.0619 \\
        DDTCDR & 0.1214 & 0.0647 & 0.1617 & 0.0898 & 0.2170 & 0.1310 & 0.1147 & 0.0638 \\
        PPGN & 0.1800 & 0.1054 & 0.2040 & 0.1109 & 0.2185 & 0.1236 & 0.1298 & 0.0688 \\
        BiTGCF & 0.1863 & 0.1011 & 0.2110 & 0.1125 & 0.2161 & 0.1225 & 0.1311 & 0.0680 \\ \midrule
        DisenCDR & \uline{0.2017} & \uline{0.1180} & \uline{0.2355} & \uline{0.1297} & \uline{0.2371} & \uline{0.1356} & \textbf{0.1513} & \textbf{0.0837} \\
        DIDA-CDR & 0.1890 & 0.1080 & 0.2021 & 0.1148 & 0.2253 & 0.1336 & 0.1342 & 0.0736 \\ 
        ETL & 0.1771 & 0.1006 & 0.2113 & 0.1162 & 0.2339 & 0.1390 & \uline{0.1407} & 0.0777 \\
        UniCDR & 0.1655 & 0.1013 & 0.1933 & 0.1100 & 0.2092 & 0.1215 & 0.1105 & 0.0603 \\ \midrule
        CoPD & \textbf{0.2182} & \textbf{0.1283} & \textbf{0.2396} & \textbf{0.1342} & \textbf{0.2606} & \textbf{0.1577} & 0.1390 & \uline{0.0790} \\  \midrule
        \bottomrule
    \end{tabular}
}
\end{table}

\subsection{Baseline Methods}
To demonstrate effectiveness, we choose 14 state-of-the-art baseline models for comparison with our proposed CoPD. These 14 baseline models can be divided into four categories, i.e., (1) Single-Domain Recommendation (NeuMF, NGCF, LightGCN), (2) Disentanglement-based Single-Domain Recommendation (DGCF, DICE), (3) Cross-Domain Recommendation (CDFM, CoNet, DDTCDR, PPGN, BiTGCF), (4) Disentanglement-Based Cross-Domain Recommendation (DisenCDR, DIDA-CDR, ETL, UniCDR).

(1) Single-domain approaches 
\begin{itemize}
    \item NeuMF~\cite{NeuralCF}: NeuMF develops techniques based on neural networks to tackle the collaborative filtering problem in recommendation.
    \item NGCF~\cite{NGCF}: NGCF stacks multiple user-item graph propagation layers to aggregate the high-order neighboring information and learn a better recommendation representation.
    \item LightGCN~\cite{LightGCN}: LightGCN is a simple design of NGCF, it removes the non-linear activation and the transformation in NGCF, exhibiting substantial improvements over NGCF.
\end{itemize}
    
(2) Disentanglement-based Single-Domain approaches 
\begin{itemize}
    \item DGCF~\cite{wang2020disentangled}: DGCF disentangles user intents on adopting the items by modeling the intent-aware interaction graphs and encouraging independence of different intents to yield disentangled representations.
    \item DICE~\cite{DICE}: DICE presents a general framework that learns structurally disentangled interest and conformity representations. We use LightGCN as the backbone of DICE in our comparison. 
\end{itemize}

(3) Cross-Domain approaches 
\begin{itemize}
    \item CDFM~\cite{CDFM}: CDFM enhances recommendations in a target domain by applying factorization machines, utilizing user interaction data from auxiliary domains as contextual information.
    \item CoNet~\cite{CoNet}: CoNet employs a customized cross-stitch neural network. Specifically, it builds deep cross-connections to enhance the prediction of user-item interactions across diverse domains.
    \item DDTCDR~\cite{DDTCDR}: DDTCDR incorporates dual learning into CDR and puts forth a deep dual transfer network. This model needs user features and item features of identical dimensions as input, acquired through matrix factorization.
    \item PPGN~\cite{PPGN}: PPGN formulates a Cross-Domain Preference Matrix to comprehensively represent interactions across diverse domains.
    \item BiTGCF~\cite{BiTGCF}: BiTGCF builds upon the Graph Collaborative Filtering network as its foundation. It enables a two-way knowledge transfer across two domains by utilizing common users as bridges. 
\end{itemize}

(4) Disentanglement-based Cross-domain approaches 
\begin{itemize}
    \item DisenCDR~\cite{DisenCDR}: DisenCDR introduces two mutual-information-based disentanglement regularizers to separate domain-shared and domain-specific information.
    \item DIDA-CDR~\cite{DIDA-CDR}: DIDA-CDR introduces an interpolative data augmentation approach to generate augmented user representations and proposes an effective method to disentangle domain-specific and domain-independent information.
    \item ETL~\cite{chen2023toward}: ETL models the joint distribution of user behaviors across domains to capture both overlapping and domain-specific features, enhancing cross-domain recommendation accuracy.
    \item UniCDR~\cite{UniCDR}: UniCDR employs domain-shared and domain-specific representations enhanced by masking mechanisms and contrastive learning to improve cross-domain recommendation performance.
\end{itemize}

\subsection{Implementation Details}
We implement our CoPD method on the PyTorch platform with Python 3.9.0. For all baseline methods and our CoPD, we use the common hyper-parameters settings reported in DisenCDR~\cite{DisenCDR}: the initializing embedding dimension is fixed as 128, the embedding parameters are initialized using the Xavier method~\cite{Xavier}, the mini-batch size is fixed as 1024, the learning rate is fixed as 0.001, the dropout rate is fixed as 0.3, the negative sampling number is fixed as 1. Additionally, we select the best hyper-parameters of the L2 regularization coefficient from \{1e-5, 1e-4\} and the number of graph encoder layers from 1 to 4.
Since the experimental settings of our CoPD are the same as DisenCDR, we directly use best results reported in DisenCDR for single-domain methods (NeuMF, NGCF and LightGCN), cross-domain-methods (CDFM, CoNet, DDTCDR, PPGN and BiTGCF) and DisenCDR. For disentanglement-based single-domain approaches (DGCF and DICE), we refer to the implementation in official code and RecBole~\cite{recbole[1.0], recbole[1.2.0], recbole[2.0]} and adjust hyper-parameters according to the optimal configurations outlined in their original papers. For DIDA-CDR, we implement it by ourselves and adjust the parameters \(\mu_1\) and \(\mu_2\) and other parameters align with the settings in DIDA-CDR. But during training, we randomly sampled 1 unseen interaction as a negative sample following DisenCDR~\cite{DisenCDR}, in contrast to the 7 negative samples mentioned in DIDA-CDR. For ETL and UniCDR, we use the source code and fine-tuned the parameters. All baseline methods and our CoPD are trained with 100 epochs for convergence, and all parameters are optimized by Adam optimizer~\cite{Adam}.\par
For the hyper-parameters in our model. We set the number of graph encoder layers to 3 and the L2 regularization coefficient to 1e-4. To determine the optimal hyper-parameters \(\lambda_{1}\) and \(\lambda_{2}\) in Equation~\eqref{eq:final_loss}, we perform a grid search within the range of \{0.01, 0.1, 1, 10\} $\bigotimes$ \{0.01, 0.1, 1, 10\}. The temperature \(\tau\) is searched within the range of \{0.01, 0.05, 0.1, 0.15, 0.2\}. After the search, we set \(\lambda_{1}=1\) and \(\lambda_{2}=1\) for Elec \& Phone, Sport \& Cloth, Douban Movie \& Book and Douban Movie \& Music, and \(\lambda_{1}=0.01\) and \(\lambda_{2}=1\) for Elec \& Cloth and Sport \& Phone. Additionally, we set \(\tau=0.05\) for all datasets. 

\begin{table}[htbp]
\renewcommand{\arraystretch}{1.3}
\centering
\caption{The overall performance comparison of CoPD and baselines on Douban CDR tasks, with the best performance highlighted in bold and the second-best performance underlined.}
\label{tab:experiment-douban}
\resizebox{\textwidth}{!}{
    \begin{tabular}{ccc|cc|cc|cc}
        \toprule
        \midrule 
        \multirow{2}{*}{Douban tasks} 
            & \multicolumn{2}{c}{Douban-Movie} & \multicolumn{2}{c|}{Douban-Book} & \multicolumn{2}{c}{Douban-Movie} & \multicolumn{2}{c}{Douban-Music} \\ \cmidrule{2-9} 
            & HR@10 & NDCG@10 & HR@10 & NDCG@10 & HR@10 & NDCG@10 & HR@10 & NDCG@10 \\ \midrule
        NGCF & 0.1026 & 0.0537 & 0.0731 & 0.0408 & 0.0916 & 0.0423 & 0.0611 & 0.0387 \\
        LightGCN & 0.1053 & 0.0549 & 0.0735 & 0.0415 & 0.0924 & 0.0425 & 0.0636 & 0.0399 \\ \midrule
        DGCF & 0.1567 & 0.0774 & 0.1227 & 0.0682 & 0.1510 & 0.0792 & 0.0655 & 0.0317 \\
        DICE & 0.1463 & 0.0785 & 0.1227 & 0.0665 & 0.1781 & 0.0939 & 0.1252 & 0.0642 \\
        \midrule
        CoNet & 0.1505 & 0.0742 & 0.1023 & 0.0447 & 0.1360 & 0.0686 & 0.0726 & 0.0356 \\
        PPGN & 0.1203 & 0.0642 & 0.1052 & 0.0478 & 0.1009 & 0.0435 & 0.0724 & 0.0403 \\
        BiTGCF & 0.1211 & 0.0646 & 0.1058 & 0.0493 & 0.1014 & 0.0441 & 0.0732 & 0.0410 \\ \midrule
        DisenCDR & 0.1639 & 0.0870 & 0.1456 & 0.0820 & 0.1631 & 0.0848 & 0.0927 & 0.0458 \\
        DIDA-CDR & 0.1815 & 0.0949 & 0.1373 & 0.0763 & 0.1883 & 0.1008 & 0.0947 & 0.0500 \\ 
        ETL & 0.1572 & 0.0716 & 0.1284 & 0.0748 & 0.1396 & 0.0742 & 0.0830 & 0.0413 \\
        UniCDR & \uline{0.2187} & \uline{0.1191} & \uline{0.1711} & \uline{0.1019} & \uline{0.2022} & \uline{0.1118} & \uline{0.1286} & \uline{0.0664} \\ \midrule
        CoPD & \textbf{0.2206} & \textbf{0.1194} & \textbf{0.1991} & \textbf{0.1152} & \textbf{0.2220} & \textbf{0.1165} & \textbf{0.1418} & \textbf{0.0785} \\ \midrule
        \bottomrule
    \end{tabular}
}
\end{table}

\subsection{Overall Performance Comparison with Baseline Models (RQ1)}
In Table~\ref{tab:experiment-amazon} and Table~\ref{tab:experiment-douban}, we provide a performance comparison of CoPD with various approaches, using metrics such as HR@10 and NDCG@10. The results indicate that the performance of CoPD significantly outperforms all competing baselines, demonstrating the effectiveness of our CoPD in solving the data sparsity problem in both domains. Based on the experimental results, we have the following observations: (1) Compared with single-domain methods, CoPD demonstrates superior performance. This indicates that our proposed global-local graph encoder framework, comprising a global graph encoder and two domain-specific local graph encoders, effectively transfers shared information across the two domains. (2) Compared with disentangle-based single-domain methods, although DGCF and DICE demonstrate outstanding performance in certain domains, outperforming several cross-domain recommendation approaches, they often yield inferior results in other domains. Our CoPD outperforms DGCF and DICE and achieves significant improvement in both domains. This indicates that CoPD produces more effective domain-shared disentanglement representations. These representations capture user-shared interest and conformity preferences and make better recommendations. (3) Compared with cross-domain methods, DisenCDR, DIDA-CDR, ETL, UniCDR and our CoPD outperform all other cross-domain methods, indicating that disentanglement-based cross-domain recommendation methods can effectively extract users’ shared information and mitigate the negative transfer problem~\cite{Zhou2023FeatureDF, Zhu2021AUF}. Disentanglement-based methods achieve excellent recommendation results by transferring this shared information across different domains. (4) Compared with disentanglement-based CDR. For DisenCDR and ETL, our proposed model not only employs domain-shared user embedding to derive shared user preference but also introduces the coherence constraints to capture the domain-shared item attributes, resulting in improved recommendations. Particularly in CDR tasks with high domain similarity in Table~\ref{tab:experiment-amazon}, we achieve significant performance improvement compared with DisenCDR and ETL. For DIDA-CDR, which disentangles user-shared embedding into domain-shared and domain-independent embeddings, our model adopts a different disentanglement approach. CoPD disentangles user-shared embedding into conformity and interest embeddings and outperforms DIDA-CDR significantly. UniCDR employs contrastive learning to maximize mutual information between specific and shared representations, which may disrupt the original distribution of domain-specific features. CoPD apply a stop gradient mechanism when maximize mutual information between specific and shared representations, which freezes the parameters of domain-specific embeddings. This ensures that domain-specific features remain unaffected and retain their uniqueness. Therefore, CoPD achieve significant performance improvement compared with UniCDR in all CDR tasks.

\subsection{Ablation Study (RQ2 \& RQ3)} 
\begin{figure}[htbp]
    \centering
    \includegraphics[width=\textwidth]{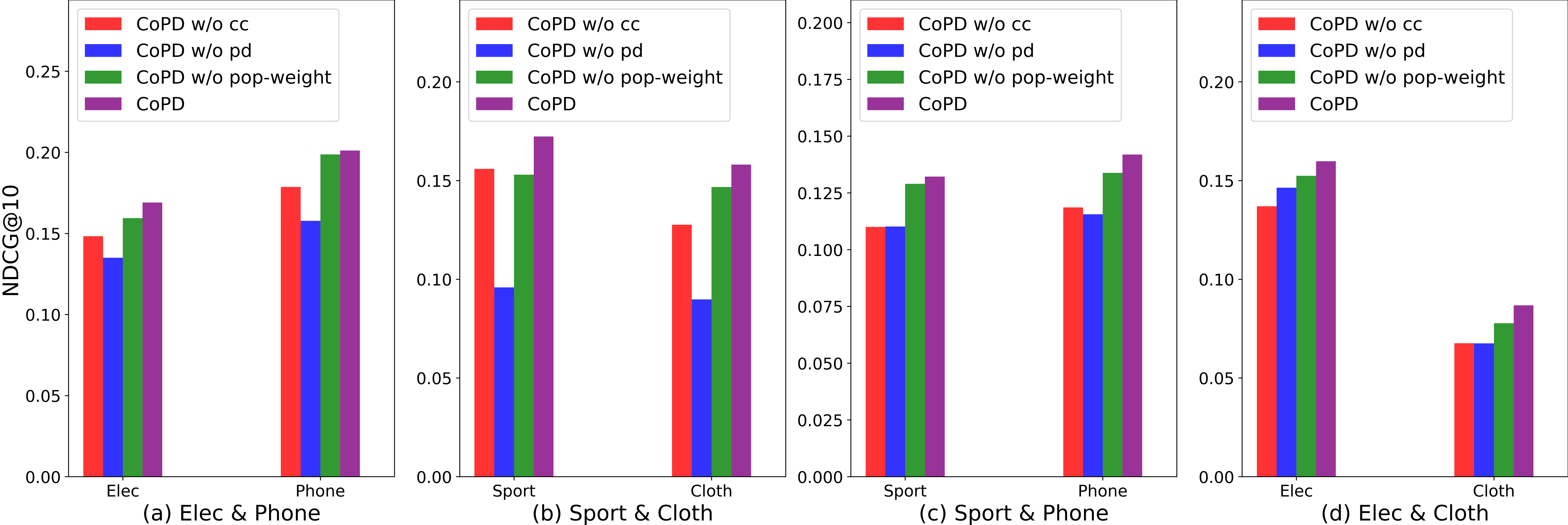}
    \caption{Performance comparison between CoPD and its two variants CoPD w/o pd and CoPD w/o cc.}
    \label{fig_ablation}
\end{figure}

\begin{figure}[htbp]
    \centering
    \includegraphics[width=\textwidth]{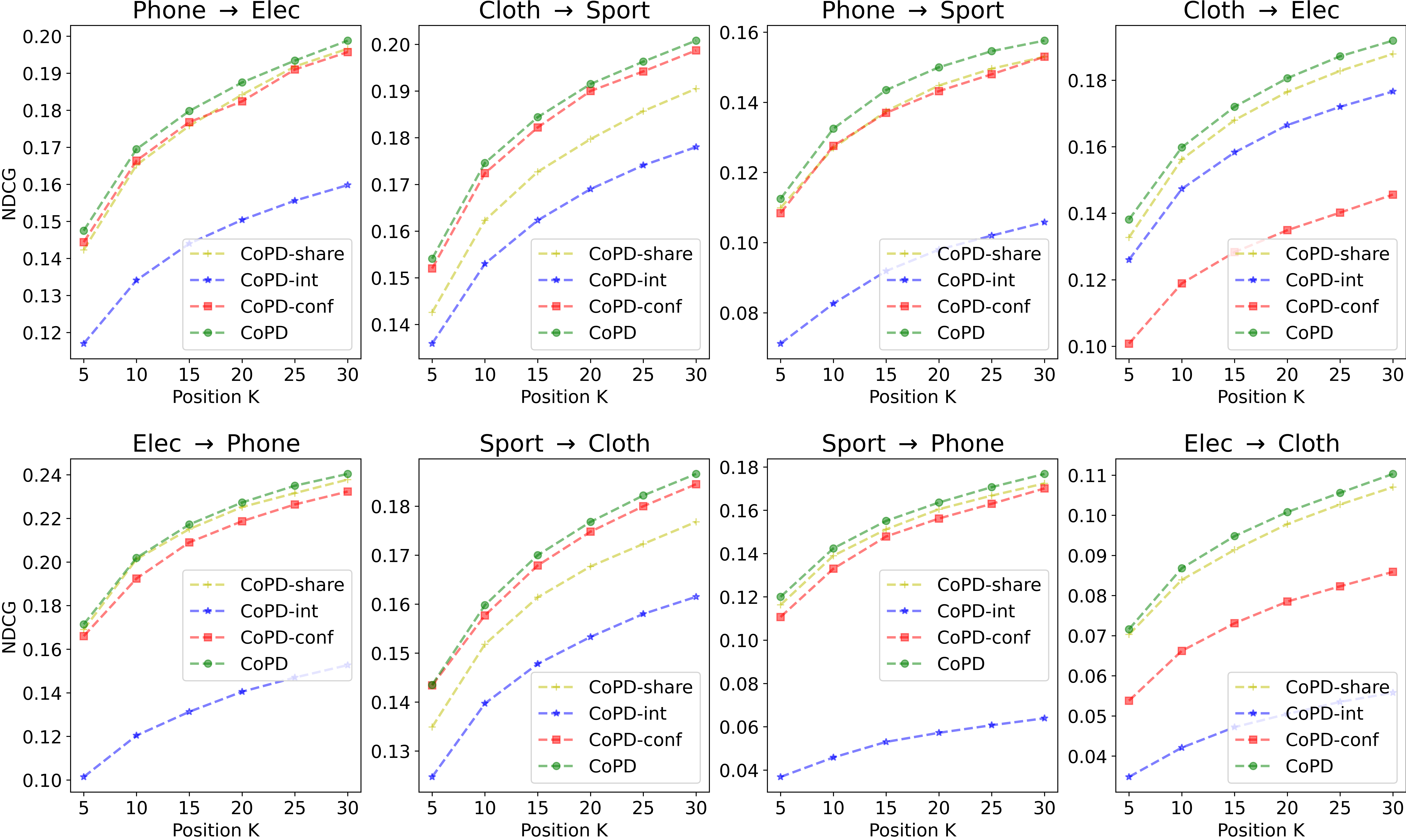}
    \caption{Effectiveness of interest embedding CoPD-int and conformity embedding CoPD-conf.}
    \label{fig_int_pop}
\end{figure}

In this subsection, we examine the effectiveness of the two modules: Coherence-guided Item Attribute Alignment (introduced in Section~\ref{sec:item}) and User Preference Disentanglement with Aligned Item Attribute (introduced in Section~\ref{sec:user}). We conduct the experiments on four variants of CoPD and show the results in Figure~\ref{fig_ablation} and Figure~\ref{fig_int_pop}.

From Figure~\ref{fig_ablation}, we investigate the effectiveness of the two modules: Coherence-guided Item Attribute Alignment and User Preference Disentanglement guided by Aligned Item Attribute. \par
\begin{itemize}
    \item CoPD w/o cc: the CoPD model without Coherence-guided Item Attribute Alignment module mentioned in Section~\ref{sec:item}. Specifically, we don't generate the domain-shared embeddings to facilitate the learning of disentangled user preference embeddings.
    \item CoPD w/o pd: the CoPD model without preference disentanglement module. Specifically, we don't disentangle the domain-shared user embedding into conformity and interest embeddings. Instead, we directly fuse the domain-shared user embedding with domain-specific user embedding for recommendation.
    \item CoPD w/o pop-weight: We train preference embeddings with non-popularity weighted Softmax loss.
\end{itemize} \par
(1) The performance of the CoPD w/o pd variant significantly decreases. This observation emphasizes the significance of the User Preference Disentanglement with the Aligned Item Attribute module in our model. Our CoPD significantly enhances by generating diverse and robust disentangled user preference embeddings from the domain-shared user embeddings for recommendation by incorporating user preference disentanglement. Additionally, CoPD outperforms CoPD w/o pop-weight, highlighting the importance of preference disentanglement with popularity-weighted Softmax loss. \par
(2) The CoPD w/o cc variant exhibits a smaller performance decrease than the CoPD w/o pd variant on Elec \& Phone and Sport \& Cloth (two high domain similarity CDR tasks). This suggests that aligning item attributes is less significant than disentangling user preferences. This observation indicates that it is possible for conventional models to capture domain-shared item attributes when users interact with items with two similar domains. In CDR tasks with lower domain similarity, such as Sport \& Phone and Elec \& Cloth, where most item attributes in these domains are distinct. It becomes challenging for recommendation models to extract domain-shared item attributes. Therefore, by employing item alignment with coherence constraints in these recommendation tasks, CoPD filters out domain-specific information and extracts more domain-shared item attributes for domain-shared item embeddings, leading to superior improvements.\par

Next, we investigate the influence of user interest and conformity embeddings on CoPD performance and the results are depicted in Figure~\ref{fig_int_pop}. \par
\begin{itemize}
    \item CoPD-share: a baseline model that lacks both interest and conformity mechanism and uses domain-shared embeddings for recommendation. We use non-weighted Softmax loss for training.
    \item CoPD-int: we don't use conformity embedding. Instead, we directly fuse the interest embedding with domain-specific user embedding for recommendation.
    \item CoPD-conf: we don't use interest embedding. Instead, we directly fuse the conformity embedding with domain-specific user embedding for recommendation.
\end{itemize} \par
(1) For these four recommendation tasks, users tend to click and give high ratings to items based on their conformity embeddings rather than interest embeddings. This indicates that conformity plays a significant role in cross-domain recommendation. Additionally, users tend to prefer popular items because a substantial portion of popular items is associated with high quality~\cite{zhao2022popularity}. Moreover, many users do not exhibit specific interest preferences in certain domains. For instance, when browsing items in the Sports domain (Accessories, Sports, and Outdoors), users prefer items with good quality and reputation but don't care about particular attributes.
\par 
(2) The user interest embedding significantly impacts the Elec domain of the Elec \& Cloth task. This illustrates that user interest preferences are also crucial, particularly when users click on items with their real interests. Additionally, we find that the Coherence-guided Item Attribute Alignment module is more effective in the Elec domain of  Elec \& Cloth task, as indicated in Figure~\ref{fig_ablation}. It indicates that extracting common item attributes aids CoPD in learning the interest embeddings of users, especially when users interact with items with their interests. \par
(3) CoPD-share outperforms CoPD-conf and CoPD-int in most CDR tasks, indicating that incorporating both conformity and interest information enhances the recommendation performance. However, CoPD surpass CoPD-share in all tasks, underscoring the effectiveness of preference disentanglement for interest and conformity.

\subsection{Analysis of Preference Disentanglement (RQ4)}
This section confirms that our CoPD can learn domain-shared user interest and conformity embeddings. We use $Conf-emb$ (blue) and $Int-emb$ (orange) to represent the learned shared conformity and interest embeddings, respectively. We use $A-emb$ (red) and $B-emb$ (green) to represent the domain-specific embeddings in domains A and B, respectively.

\begin{figure}[htbp]
    \centering
    \includegraphics[width=0.7\linewidth]{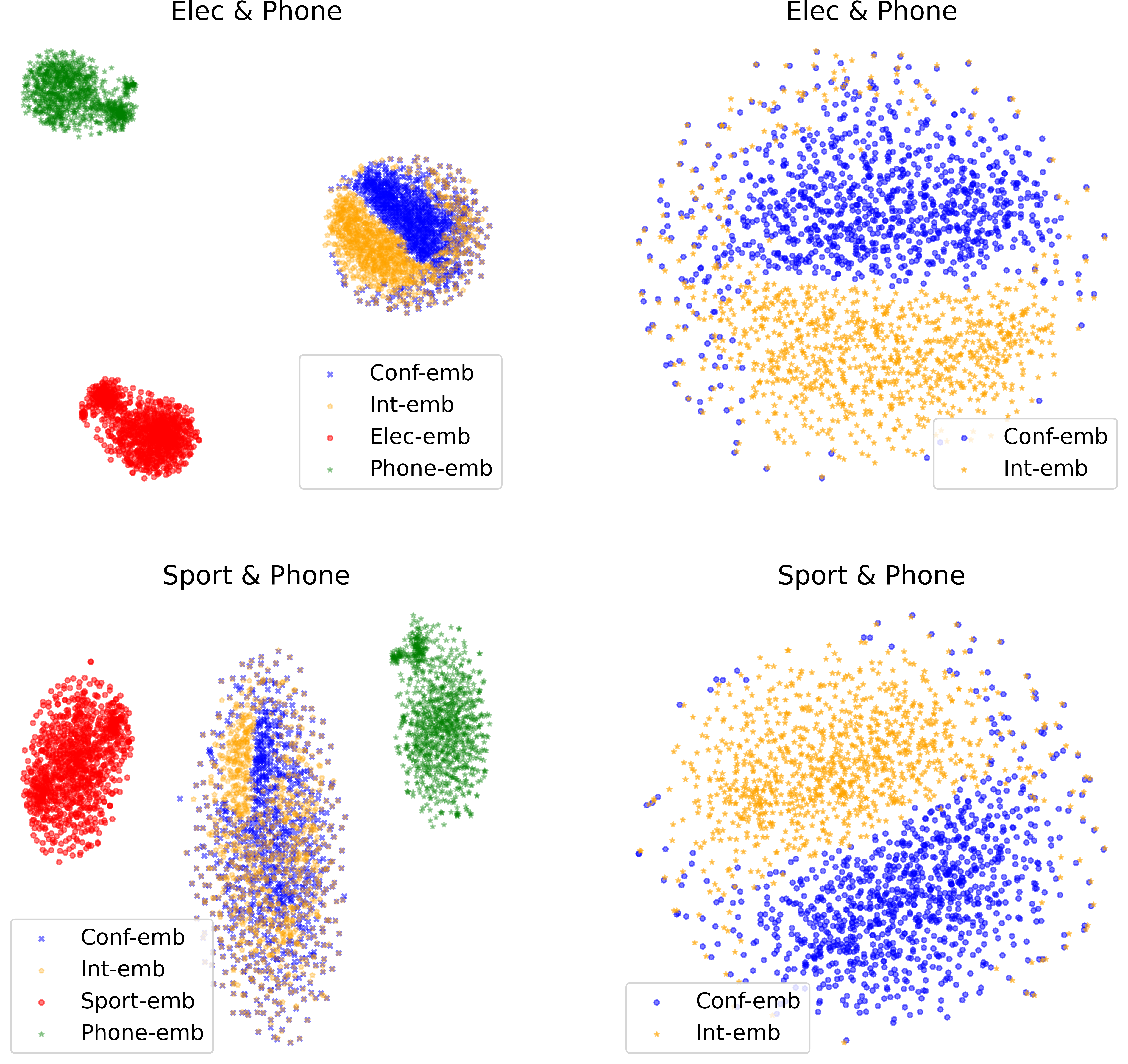}
    \caption{Visualization of conformity embeddings (blue), interest embeddings (orange), domain-A-specific embeddings (red) and domain-B-specific embeddings (green).}
    \label{fig:t-SNE}
\end{figure}

We visualize the learned embeddings using t-SNE~\cite{t-SNE}, as shown in Figure~\ref{fig:t-SNE}. In the left columns, the two-dimensional embeddings of domain-shared and domain-specific are distinctly separated, indicating that our CoPD effectively disentangles user preference embeddings into domain-shared and domain-specific embeddings. However, interest and conformity embeddings are not clearly distinguishable due to the confounding influence of domain-specific embeddings. In the right columns, we exclusively train domain-shared interest and conformity embeddings in T-SNE. The conformity embeddings and interest embeddings are distinguishable, demonstrating the effectiveness of the disentanglement by popularity-weighted Softmax loss.
In conclusion, unlike previous approaches that only rely on domain-shared user embeddings, our disentanglement method disentangles the domain-shared user embeddings into interest and conformity embeddings, resulting in improved performance.
\begin{figure}[htbp]
    \centering
    \includegraphics[width=0.7\linewidth]{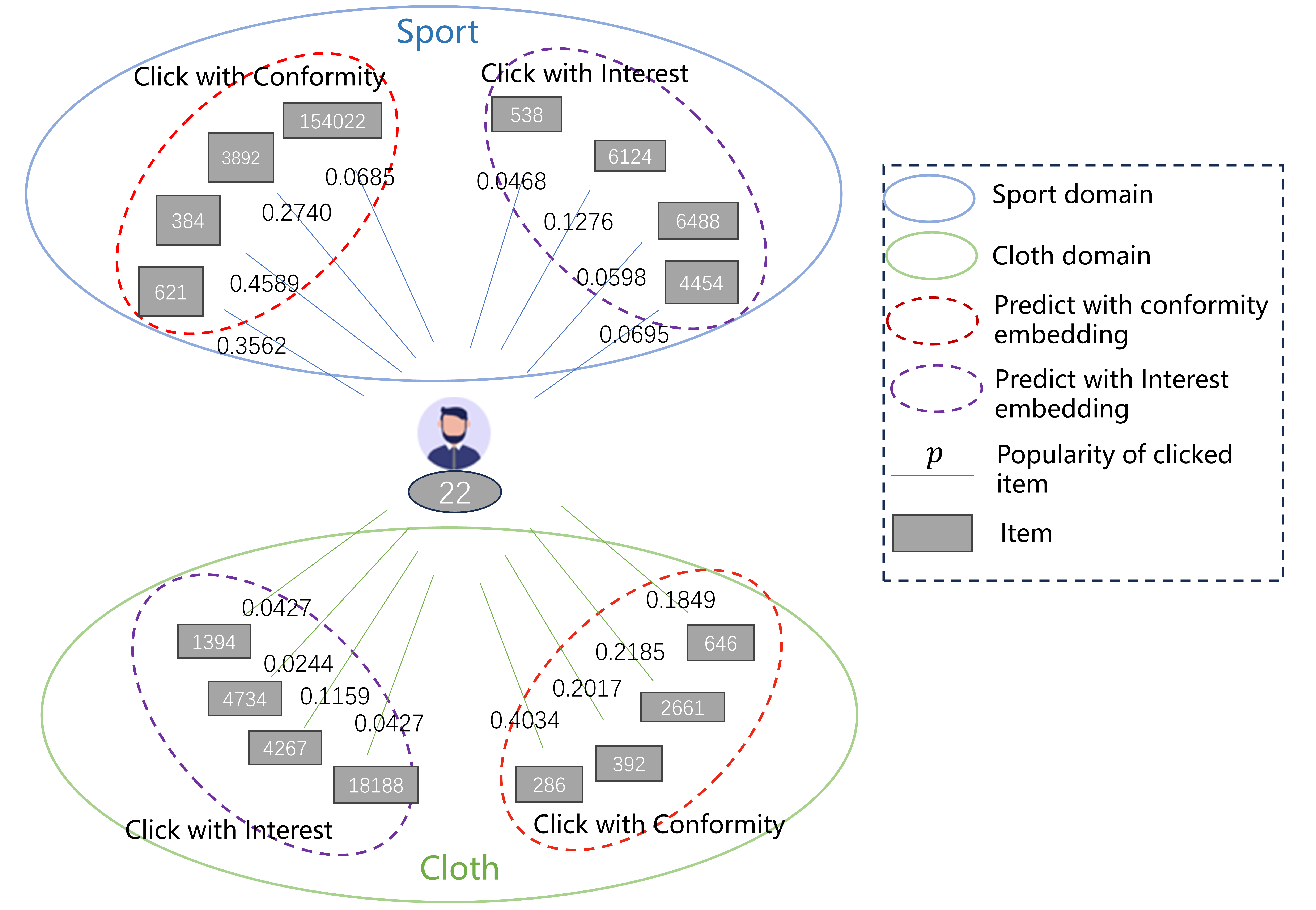}
    \caption{The case study of user \#22 on Sport \& Cloth task.}
    \label{fig_case}
\end{figure}

\subsection{Case Study (RQ4)}
In this section, we present a case study to show the effectiveness of our model in capturing user interest and conformity preferences, as illustrated in Figure~\ref{fig_case}. We select the top-4 items in each domain with different preference recommendations in the Sport \& Cloth task. In observing the recommendation made by the user conformity embedding (red dotted circle), a clear trend emerges: items with high ratings also exhibit high popularity. This implies that the user \#22, with his conformity preference, tends to click on popular items. 
For example, in the Sport domain, item \#384 has a popularity of 0.4589; in the Cloth domain, item \#286 has a popularity of 0.4034. 
Furthermore, in observing the recommendation made by the user interest embedding (purple dotted circle), the popularity of clicked items remains consistently low. This indicates that these items are popularity-irrelevant and reflect the user's real interest. 
For instance, in the Sport domain, item \#538 has a popularity of 0.0468, and in the Cloth domain, item \#4734 has a popularity of 0.0244. 
When disentangling user preferences using popularity-weighted loss, it's worth noting that our CoPD successfully achieves the desired separation between user conformity and interest preferences.
\begin{figure}[htbp]
    \centering
    \includegraphics[width=\linewidth]{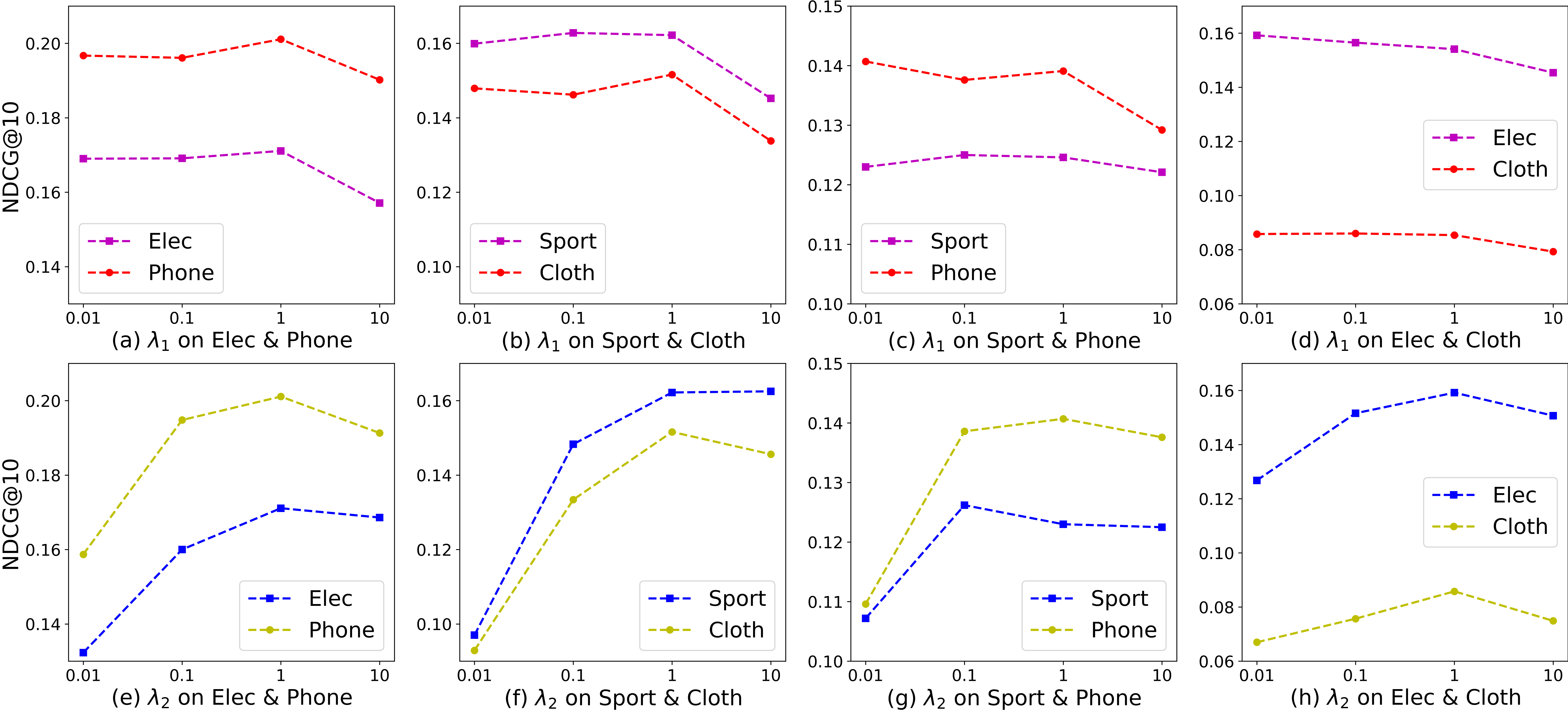}
    \caption{(a)-(d): Impact of $\lambda_{1}$. (e)-(h): Impact of $\lambda_{2}$.}
    \label{fig_lamda}
\end{figure}

\begin{figure}[htbp]
    \centering
    \includegraphics[width=\linewidth]{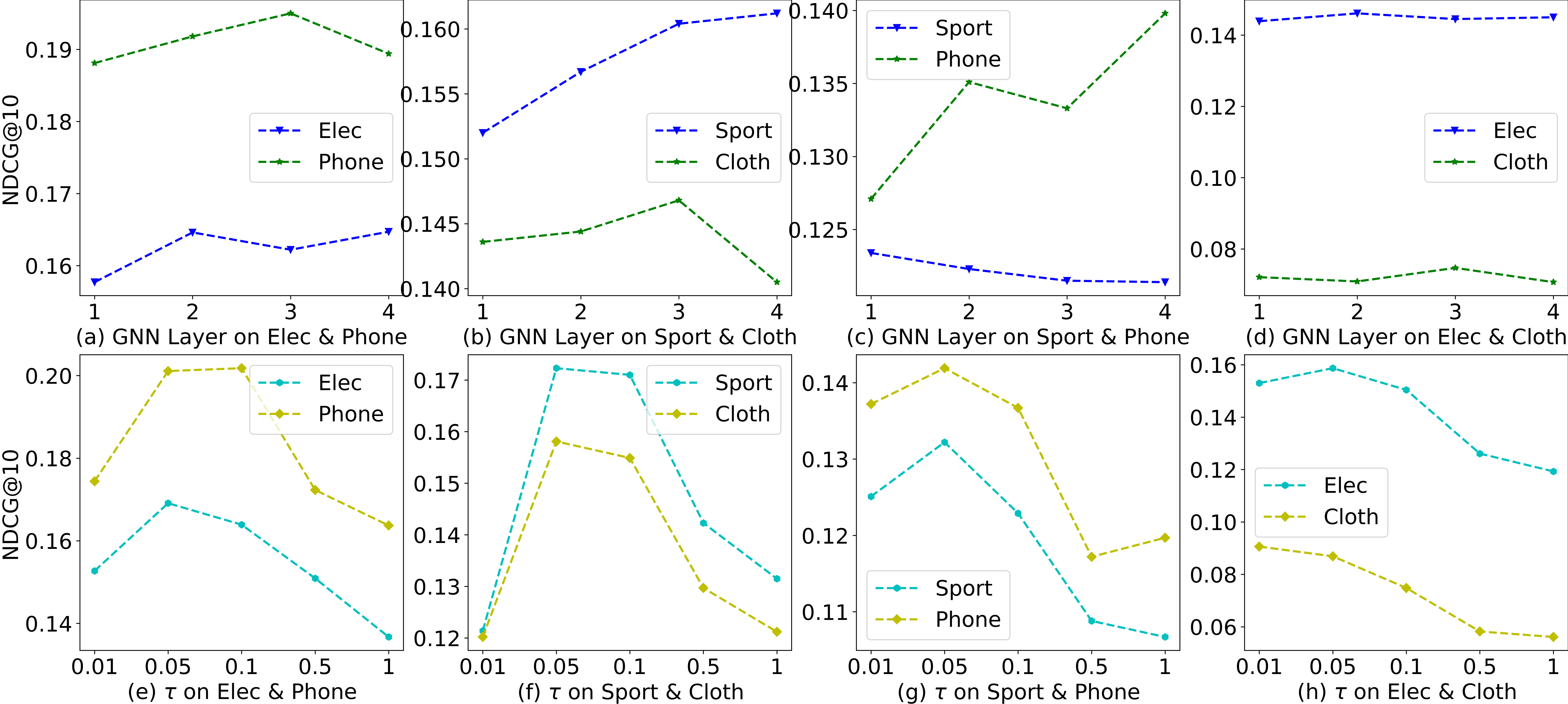}
    \caption{(a)-(d): Impact of the number of GNN layers. (e)-(h): Impact of \(\tau\).}
    \label{fig_l_temp}
\end{figure}

\subsection{The Impact of Hyper-parameters (RQ5)}
In this subsection, (1) we initially test the sensitivity of two hyper-parameters \(\lambda_{1}\) and \(\lambda_{2}\) in Equation~\eqref{eq:final_loss}. According to Figure~\ref{fig_lamda}, we conclude that \(\lambda_{2}\) is more sensitive than \(\lambda_{1}\). $\lambda_1$ represents a trade-off of coherence constraints to synchronize domain-shared embeddings with domain-specific ones. A higher $\lambda_1$ reduces redundant information in domain-shared embeddings but risks losing valuable domain-shared information. Conversely, a lower $\lambda_1$  preserves domain-shared information more effectively, although it may not eliminate redundancy as efficiently. From Figure~\ref{fig_lamda} (a)-(d), we observe that a slightly higher $\lambda_1$ (e.g., 1) improve the performance in tasks with high domain similarity by reducing redundant information (Figure~\ref{fig_lamda}~a and \ref{fig_lamda}~b), while a lower $\lambda_{1}$ (e.g., 0.01) preserve the domain-shared information and performs better in tasks with low domain similarity (Figure~\ref{fig_lamda}~c and \ref{fig_lamda}~d). However, when $\lambda_1$ is set too high (e.g., 10), performance declines sharply due to the excessive loss of shared information and the increased influence of domain-specific information.
For \(\lambda_2\), a slightly larger value, such as 1, yields the best performance. Therefore, with a larger value of \(\lambda_2\), CoPD can pay more attention to disentangling user interest and conformity embeddings. However, a larger value of \(\lambda_2\) reduces performance, as it decreases the influence of domain-specific information. 
(2) Next, we conducted experiments to investigate how the number of GNN layers influences performance and report the results in Figure~\ref{fig_l_temp} (a)-(d). The 1-layer and 4-layer variants achieved the worst results because the 1-layer GNN can't aggregate enough information, and the 4-layer variant suffers from the over-smoothing problem. Specifically, the best number of GNN layers on Sport is 1 in Figure~\ref{fig_l_temp} (c). This is due to the fact that the number of GNN layers influence both the global graph $G^{S}$ and single-domain graph $G^{A}$ (the Sport graph). As the GNN layers increase, the Sport domain receive more redundant information from Phone domain via the global graph $G^{S}$. (3) Finally, as illustrated in Figure~\ref{fig_l_temp} (e)-(h), the temperature \(\tau\) is crucial for model performance. We observe that a large value of \(\tau\) leads to poor performance, aligning with experimental results reported in \cite{NCL}.

\section{CONCLUSION}
This paper introduces a Coherence-guided Preference Disentanglement (CoPD) method to tackle challenges arising from data sparsity in recommendation tasks. The CoPD framework comprises three components. The first component is coherence-guided item attribute alignment, where a global-local graph encoder framework and coherence constraints are introduced to acquire common item attributes. The global-local graph encoder framework facilitates the effective learning of domain-shared and domain-specific embeddings, and coherence constraints preserve common and essential attributes in domain-shared embeddings while eliminating redundant attributes from another domain. \par
The second component is user preference disentanglement with aligned item attributes. For learning disentangled user preferences, we utilize the popularity-weighted Softmax loss between aligned item embeddings and domain-shared user embeddings, guiding the model to emphasize the user's conformity preferences toward common attributes of popular items and directing the user's interest preferences toward the common attributes of popular-irrelevant items. The third component is the fusion of domain-specific and domain-shared embeddings, aimed at integrating these embeddings into comprehensive user preferences and item attributes. Experimental results demonstrate the superior performance of CoPD over state-of-the-art baselines. \par
In the future, we plan to enhance our disentanglement techniques by integrating side information, such as user reviews and item category data, to more effectively capture users' shared preferences across domains and enhance cross-domain recommendations' performance. Furthermore, we plan to devise specific coherence constraints for different cross-domain recommendation tasks, such as cold-start user and cold-start item recommendations. Finally, we will investigate the preference disentanglement in group-based recommendations. Motivated by the insights of this paper, we hold that users within the same group share common preferences, referred to as conformity and interest studied in this paper. By disentangling users' group preferences into conformity and interest, we anticipate enhancing the performance of group-based recommendations. 

\section*{Acknowledgment}
This work is supported by the National Natural Science Foundation of China (grant No. 62002052). The Shenzhen Longhua Fundamental Research Program (No. 10162A20230325B73A546). The Start-up funding provided by Charles Darwin University. The Australian Research Council partially supports this work under the streams of Future Fellowship (Grant No. FT210100624), the Discovery Project (Grant No. DP240101108), and the Linkage Project (Grant No. LP230200892).

\bibliographystyle{ACM-Reference-Format}
\bibliography{ref} 

\end{document}